\documentclass{aa} 

\newcommand{\CII}{[C\,{\scriptsize II}]}

\newcommand{\OI}{[O\,{\scriptsize I}]}

\newcommand{\CI}{[C\,{\scriptsize I}]}

\usepackage{graphicx}
\usepackage{txfonts}
\usepackage{hyperref}
\usepackage{natbib}
\begin{document} 

   \title{Dense gas formation in the Musca filament due to the dissipation of a supersonic converging flow}


   \author{L. Bonne 
          \inst{1}
          \and
          N. Schneider\inst{2}
          \and
          S. Bontemps\inst{1}
          \and
          S. D. Clarke \inst{2}
          \and
          A. Gusdorf\inst{3,4}
          \and
          A. Lehmann\inst{3}
          \and
          M. Steinke \inst{2} 
          \and
          T. Csengeri \inst{1,5}
          \and
          S. Kabanovic \inst{2} 
          \and
          R. Simon \inst{2}  
          \and
          C. Buchbender\inst{2}
          \and
          R. G\"usten\inst{5}
          }

   \institute{Laboratoire d'Astrophysique de Bordeaux, Universit\'e de Bordeaux, CNRS, B18N,
              all\'ee Geoffrey Saint-Hilaire, 33615 Pessac, France\\
              \email{lars.bonne@u-bordeaux.fr}
         \and
             I. Physikalisches Institut, Universit\"at zu K\"oln, Z\"ulpicher Str. 77, 50937 K\"oln, Germany
         \and
             Laboratoire de Physique de l'\'Ecole Normale Sup\'erieure, ENS, Universit\'e PSL, CNRS Sorbonne Universit\'e de Paris, Paris, France
         \and
             Observatoire de Paris, PSL University, Sorbonne Universit\'e, LERMA, 75014 Paris, France   
         \and
             Max-Planck-Institut f\"ur Radioastronomie, Auf dem H\"ugel 69, 53121 Bonn, Germany\\
             }

   \date{Draft of \today}
   
 
  \abstract
      {Observations with the \textit{Herschel} Space Telescope have
        established that most star forming gas is organised in
        filaments, a finding that is supported by numerical simulations of the supersonic
        interstellar medium (ISM) where dense filamentary structures
        are ubiquitous. We aim to understand the formation of
          these dense structures by performing observations covering the
          $^{12}$CO(4$\to$3), $^{12}$CO(3$\to$2), and various  
          CO(2-1) isotopologue lines of the Musca filament, using the APEX
          telescope. The observed CO intensities and line ratios cannot be explained by PDR (photodissociation region) emission
        because of the low ambient far-UV field that is strongly
        constrained by the non-detections of the \CII\ line at 158
        $\mu$m and the \OI\ line at 63 $\mu$m, observed with the
        upGREAT receiver on SOFIA, as well as a weak \CI\ 609 $\mu$m
        line detected with APEX.  We propose that the 
        observations are consistent with a scenario in which shock 
        excitation gives rise to warm and dense gas close to the   
        highest column density regions in the Musca filament.  
        Using shock models, we find that the
        CO observations can be consistent with excitation by J-type
        low-velocity shocks. A qualitative comparison of the 
         observed CO spectra with
         synthetic observations of dynamic filament formation
          simulations shows a good agreement with the signature of a
        filament accretion shock that forms a cold and dense filament
        from a converging flow. The Musca filament is thus found to be 
        dense molecular post-shock gas.
        %
          Filament accretion
          shocks that dissipate the supersonic kinetic energy of
          converging flows in the ISM may thus play a prominent role
          in the evolution of cold and dense filamentary structures.  }


   \maketitle
%

\section{Introduction} \label{sec:introduction}

Observations with the \textit{Herschel} Space Telescope have revealed that filamentary
structures are ubiquitous in the supersonic interstellar medium (ISM)
\citep[e.g.][]{Andre2010,Molinari2010,Henning2010,Arzoumanian2011,Schneider2012}. However,
there is an ongoing discussion regarding the nature and diversity of the
filaments, namely, whether they are sheets viewed edge-on or, rather, dense gas
cylinders. It is also considered whether we observe a full range of filament classes: from
cross-sections of sheets to dense, star-forming cylindrical
structures. In any case, understanding the nature, formation and
evolution of filaments is essential as they are the sites of star
formation. This was demonstrated by a number of recent studies which
showed that pre- and protostellar cores are mostly located in
filaments
\citep[e.g.][]{Polychroni2013,Andre2014,Schisano2014,Konyves2015,Marsh2016,Rayner2017}.
%
In numerical simulations of the ISM, filaments are omnipresent and can
form in various ways: through shocks in (magnetic) supersonic
turbulent colliding flows
\citep[e.g.][]{Padoan2001,Jappsen2005,Smith2016,Federrath2016,Clarke2017,Inoue2018},
during the global gravitational collapse of a cloud
\citep{Gomez2014,VazquezSemadeni2019}, through velocity shear in a
magnetised medium \citep{Hennebelle2013}, or via the gravitational
instability of a sheet \citep{Nagai1998}.

However, it is challenging to find observational signatures that
reveal how a filament is formed. In the view of large-scale
  colliding flows, filament formation is associated with the
  generation of warm gas from low-velocity shocks. An observational
  signature of these shocks are anomalously bright mid- and high-J CO
  lines, that is, line integrated intensities that are higher than
  expected considering only heating from the far-ultraviolet (FUV)
  field \citep[e.g.][]{Pon2012}.  Spectroscopic {\sl Herschel}
  observations of such lines towards the Perseus and Taurus clouds detected 
  this excess emission \citep{Pon2014,Larson2015}, where it was proposed to be
  the result of low-velocity shocks (v$_{shock} <$ 4 km s$^{-1}$), 
  dissipating the overall, generic supersonic turbulence of a
  molecular cloud.
 
\begin{figure*}
\begin{center}
\includegraphics[width=0.8\hsize]{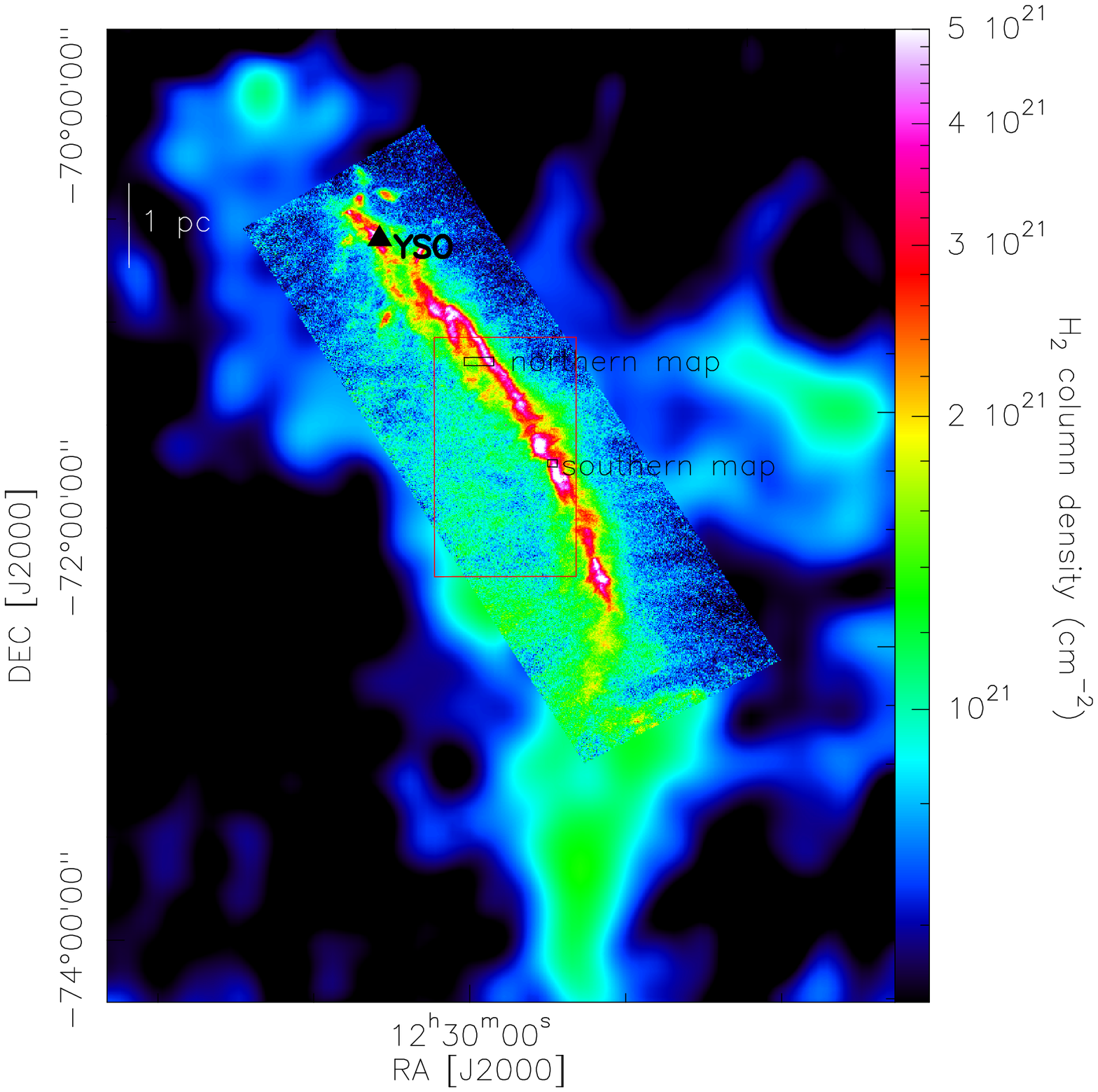}
\caption{\textit{Herschel} column density map of the Musca
  filament \citep{Cox2016} inserted into the extinction map of the
  Musca cloud \citep{Schneider2011}, which is scaled to the
  \textit{Herschel} column density map. The filament crest is defined
  as N $>$ 3$\cdot$10$^{21}$ cm$^{-2}$ (purple and white). The
    region outlined by a red rectangle is the zoom displayed in
    Fig. \ref{Map250}.  The ambient cloud is displayed in green and
  blue. The northern and southern areas mapped with FLASH+ on the APEX
  telescope are indicated in black. The triangle indicates the
  location of the only YSO in the Musca filament.}
\label{Map}
\end{center}
\end{figure*}

\begin{figure}
\includegraphics[width=\hsize]{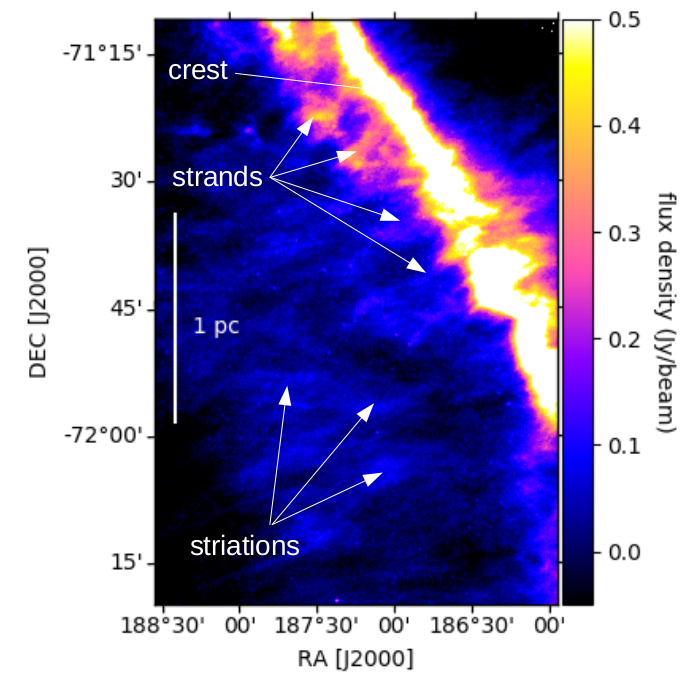}
\caption{Zoom (red box in Fig. \ref{Map}) into the \textit{Herschel}
  250 $\mu$m map of Musca, indicating the striations, strands and the
  filament crest \citep{Cox2016}.}
\label{Map250}
\end{figure}

In this paper, we report on the excess emission seen in mid-J $^{12}$CO 
lines observed around the Musca filament and propose that in low- to moderate 
density regions, exposed to a very weak FUV field, these lines may 
also serve as a tracer for low-velocity shocks. The Musca filament 
is located at a distance of 140 pc \citep{Franco1991}. 
We adopt the nomenclature from \citet{Cox2016} for the different features 
of the Musca filament and we indicate them in Fig. \ref{Map250} (see also Fig. 2 in 
\citet{Cox2016}). First, there is the high column density filament crest 
with N$>$3 10$^{21}$ cm$^{-2}$ (similar value to the one used by \citet{Cox2016} 
which is N$>$2.7 10$^{21}$ cm$^{-2}$). In addition, C$^{18}$O(2-1) emission basically disappears 
below this column density value. Attached to the filament crest are 
intermediate column density (N$\sim$2 10$^{21}$ cm$^{-2}$) hair-like structures called strands, with a size of $\sim$0.2-0.4 pc. 
Even further outwards at larger distances, narrow straight structures of up to a few parsec lengths,
called striations \citep{Goldsmith2008,Palmeirim2013,AlvesDeOliveira2014, Heyer2016,Cox2016,Malinen2016} are seen. 
They are mostly orthogonal to the crest and located in the ambient cloud. 
The ambient cloud then refers to the
environmental gas, traced by the extinction map in Fig. \ref{Map},
embedding the denser Musca filament crest and
strands. The filament crest is velocity-coherent \citep{Hacar2016} and
has only one young stellar object (YSO) located at the northern end
\citep[e.g.][Fig. \ref{Map}]{VilasBoas1994}.  It was recently proposed
that Musca is a sheet seen edge-on \citep{Tritsis2018}. However, one
of the main results of a companion paper (Bonne et al., in prep.,
hereafter Paper I), is that the Musca filament crest is more
consistent with a cylindrical geometry at a density of n$_{H_{2}}
\sim$ 10$^{4}$ cm$^{-3}$. Because other authors \citep{Cox2016,Kainulainen2016,Hacar2016} also support this cylindrical
geometry, we adopt this view as a working hypothesis.

In Paper I, the large scale kinematics and physical conditions in the
Musca filament and cloud are studied with the main CO(2-1) isotopologues,
as well as the relation between Musca and the Chamaeleon-Musca
complex. In this paper, we present $^{12}$CO(4-3) and $^{12}$CO(3-2) lines 
and argue for the presence of a blueshifted excess velocity 
component best visible towards the filament crest and strands. A detailed  
analysis of this $'$blueshifted component$'$ is the objective of this paper.\\
Here, we present a non-LTE CO excitation
analysis of the blueshifted component, as well as complementary
observations of the far-infrared fine structure transitions of carbon
(\CI), oxygen (\OI) and ionised carbon (\CII), which reveals a warm
gas component around the filament. We show that the excitation of this
warm gas can be explained by a filament accretion shock, that is, a
low-velocity shock due to mass accretion on a filament from a
converging flow in an interstellar cloud.


\section{Observations} \label{sec:observations}

\subsection{APEX}
\begin{figure}
\includegraphics[width=\hsize]{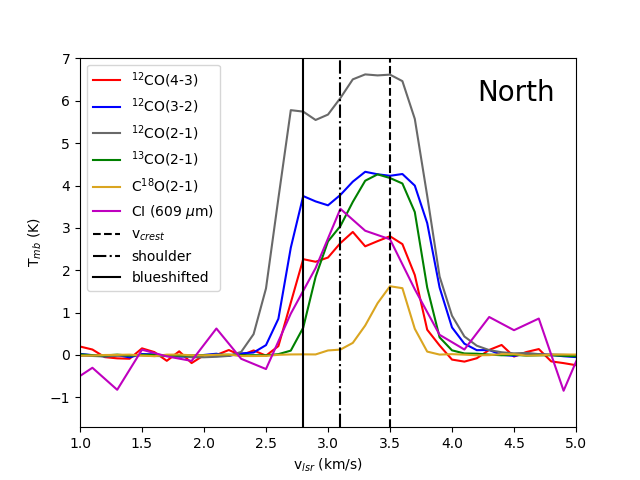}
\includegraphics[width=\hsize]{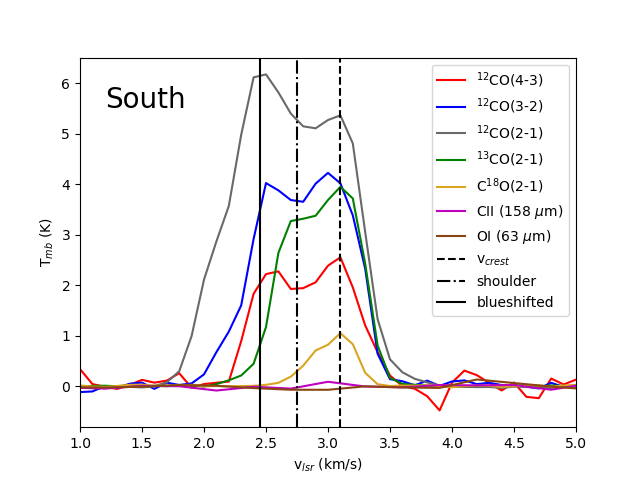}
\caption{\textbf{Top}: APEX CO and \CI\ spectra in the northern
    map, averaged over the filament crest (N $>$ 3$\cdot$10$^{21}$cm$^{-2}$).  
    \textbf{Bottom}: APEX CO spectra averaged over the
    southern map and the 7-pixel averaged SOFIA \CII\ and \OI\ lines
    that are not detected above the 3$\sigma$ level. We note the
    prominent blueshifted velocity component around 2.8 km s$^{-1}$ in
    the north and at 2.5 km s$^{-1}$ in the south, only visible in
    the $^{12}$CO lines.}
\label{spectraCrestNorth}
\end{figure}

We used the FLASH460 receiver on the APEX telescope
\citep{Guesten2006,Klein2014}, and obtained a single pointing in \CI\ $^{3}$P$_{1}$-$^{3}$P$_{0}$
at 609 $\mu$m towards the Musca filament at $\alpha_{\text{2000}}$ =
12$^{h}$28$^{m}$55$^{s}$ and $\delta_{\text{2000}}$ =
-71$^{o}$16$^{\prime}$55$^{\prime\prime}$, using the OFF position
$\alpha_{\text{2000}}$ = 12$^{h}$41$^{m}$38$^{s}$ and
$\delta_{\text{2000}}$ = -71$^{o}$11$^{\prime}$00$^{\prime\prime}$
\citep{Hacar2016}. The beamsize is 13$^{\prime\prime}$ and the
spectral resolution is $\sim$ 0.05 km s$^{-1}$. The data reduction
was done with the CLASS
software\footnote{http://www.iram.fr/IRAMFR/GILDAS}. A main beam
efficiency\footnote{http://www.apex-telescope.org/telescope/efficiency/}
$\eta_{\text{mb}}$ = 0.49 was applied for \CI, a baseline of order one was removed, and a correction for a 490 kHz
shift in the FLASH460 instrument was performed (see Paper I, F. Wyrowski, priv. comm.).  Fitting
the \CI\ single pointing (Fig.~\ref{spectraCrestNorth}) with a
Gaussian profile provides a peak temperature brightness T$_{\rm mb}$ = 3.5 K and FWHM = 0.8 km s$^{-1}$
(or 3.4$\cdot$10$^{-7}$ erg s$^{-1}$ cm$^{-2}$ sr$^{-1}$).\\
The CO observations with APEX are presented in more detail in Paper I, but here we shortly summarise 
the most important facts. All data was obtained in 2017 and 2018, using FLASH345 and FLASH460 for the 
$^{12}$CO(3-2) and $^{12}$CO(4-3) mapping of the northern and southern regions (Fig. \ref{Map}). 
The FLASH345 (FLASH460) observations have a spectral resolution of 0.033 (0.05) km s$^{-1}$ and an angular 
resolution of 18$''$ (14$''$). Main beam efficiencies of $\eta_{mb}$=0.65 (0.49) were applied to the 
antenna temperatures. We here use data sampled to 0.1 km s$^{-1}$ and smoothed to a resolution of 28$''$. 
This allows a better comparison to the CO(2-1) APEX data which were taken with the PI230 receiver. The spectral 
resolution for CO(2-1) is 0.08 km s$^{-1}$ at an angular resolution of 28$''$. Here, we applied a main beam efficiency 
of $\eta_{mb}$=0.68. The center (0,0) position of the maps is 
$\alpha_{\text{2000}}$ = 12$^{h}$28$^{m}$58$^{s}$, $\delta_{\text{2000}}$ = -71$^{\circ}$16$^{\prime}$55$^{\prime\prime}$ 
for the northern map and 
$\alpha_{\text{2000}}$ = 12$^{h}$24$^{m}$46$^{s}$, $\delta_{\text{2000}}$ = -71$^{\circ}$47$^{\prime}$20$^{\prime\prime}$ 
for the southern map.

\begin{figure*}
\includegraphics[width=1.\hsize]{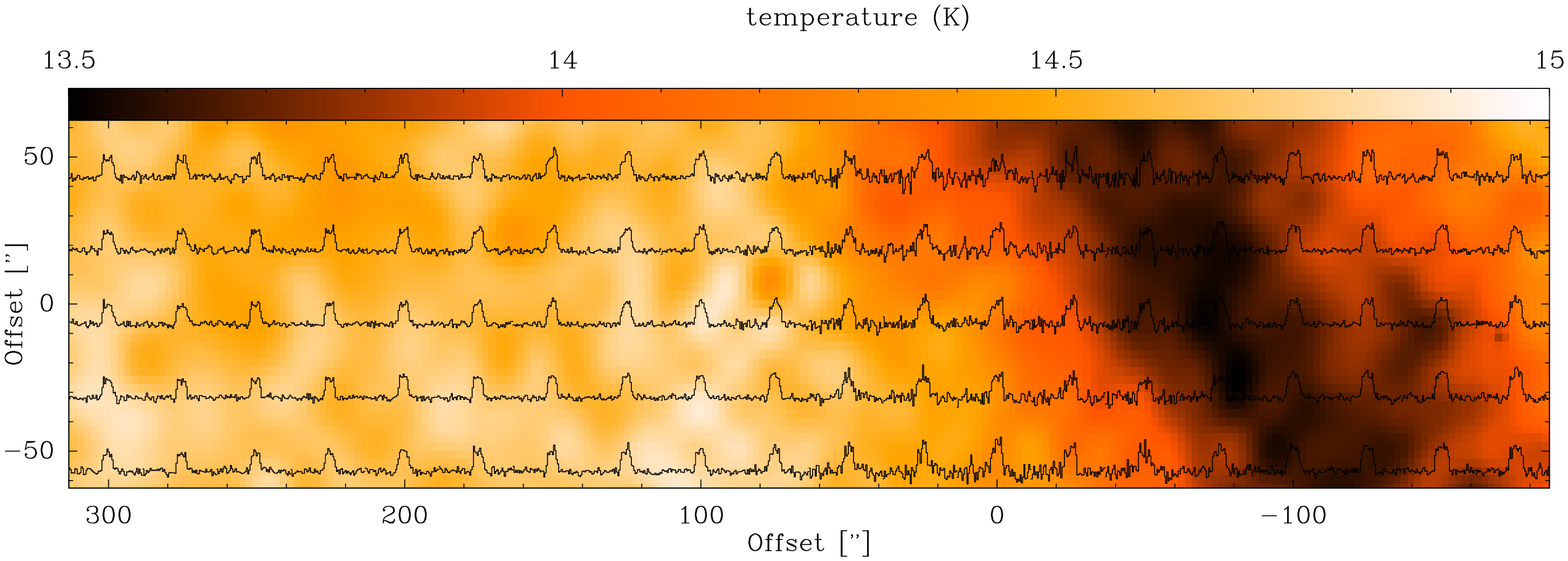}
\includegraphics[width=1.\hsize]{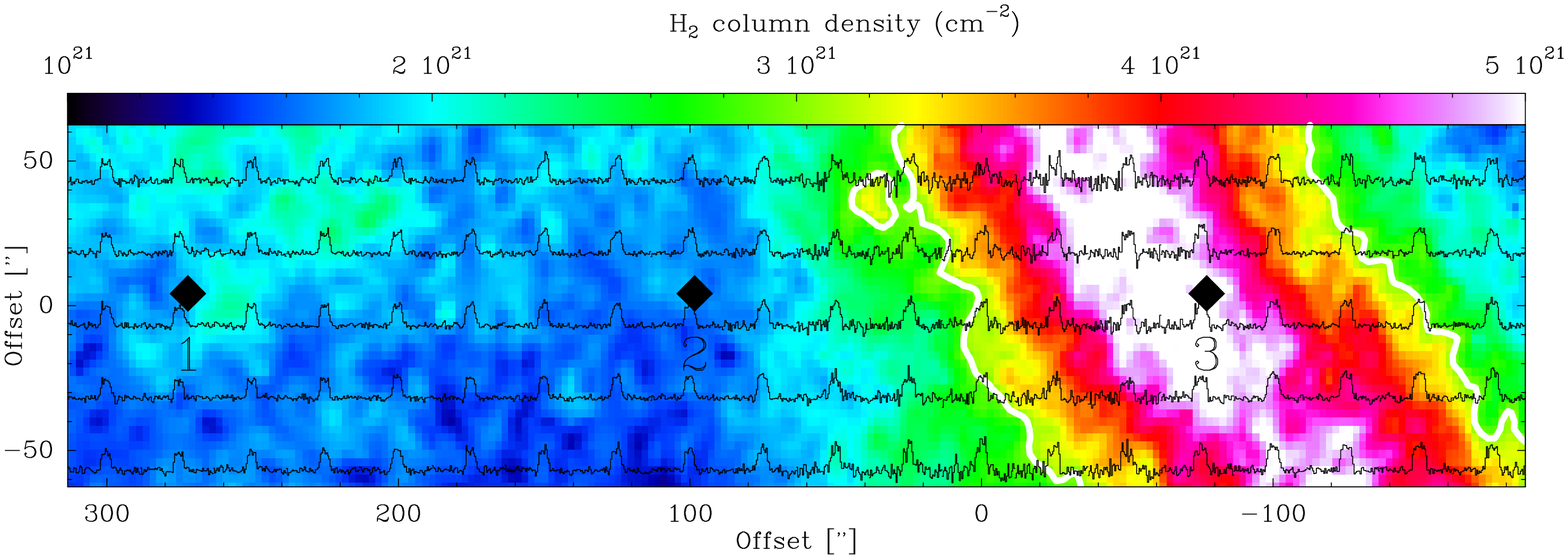}
\begin{center}
\includegraphics[width=0.325\hsize]{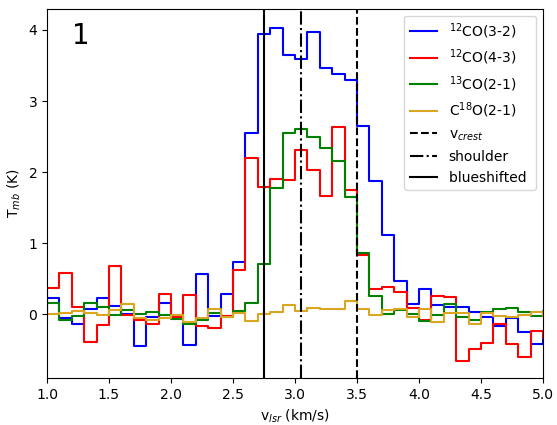}
\includegraphics[width=0.325\hsize]{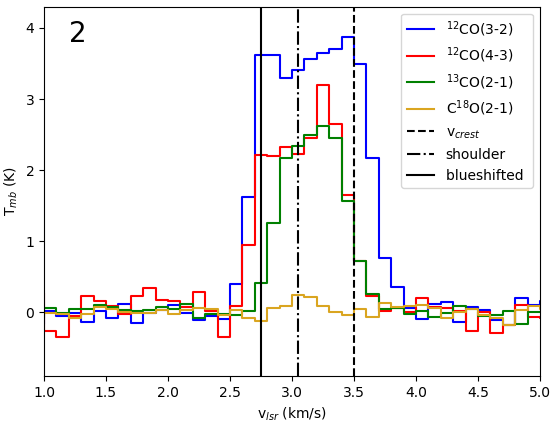}
\includegraphics[width=0.325\hsize]{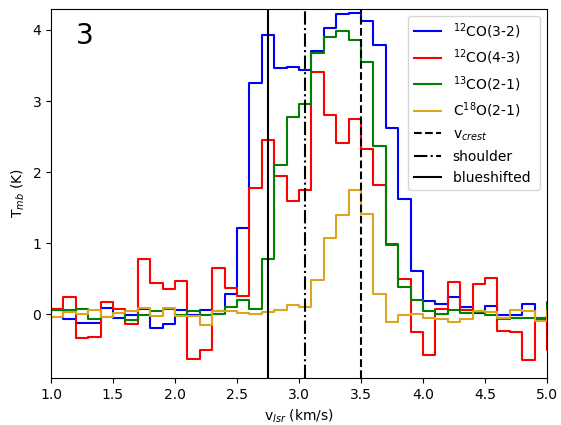}
\end{center}
\caption{\textbf{Top:} $^{12}$CO(4-3) spectra overlaid on the \textit{Herschel} dust temperature map. The (0,0) position of the map is 
$\alpha_{\text{2000}}$ = 12$^{h}$28$^{m}$58$^{s}$, $\delta_{\text{2000}}$ = -71$^{o}$16$^{\prime}$55$^{\prime\prime}$ \textbf{Middle:} $^{12}$CO(4-3) spectra overlaid on the
  northern \textit{Herschel} column density map. The white contours
  indicate Musca filament crest (N $>$ 3$\cdot$10$^{21}$
  cm$^{-2}$). The black squares indicate the locations of the spectra
  displayed below. \textbf{Below:} Zoom into spectra at selected
  positions in the map. This shows that there is $^{12}$CO(4-3) and
  $^{12}$CO(3-2) emission at the velocity of the blueshifted component
  over the full map.}
  %
\label{singleSpectraCO43}
\end{figure*}

\subsection{SOFIA}
In June 2018, the upGREAT instrument \citep{Risacher2018} on the
Stratopsheric Observatory for Far-Infrared Astronomy (SOFIA) was used
for a 70 minute single pointing of the 7 pixel array covering the
atomic oxygen \OI\ $^{3}$P$_{1}$-$^{3}$P$_{2}$ fine structure line at
63 $\mu$m and the ionised carbon \CII\ $^{2}$P$_{3/2}$-$^{2}$P$_{1/2}$
fine structure line at 158 $\mu$m towards the southern APEX map at
$\alpha_{\text{2000}}$ = 12$^{h}$24$^{m}$41$\fs$6,
$\delta_{\text{2000}}$ = -71$^{\circ}$46$^{\prime}$41$\farcs$0. These
observations were performed in the single beam switching mode with a
chop amplitude of 150$^{\prime\prime}$ towards a location with weak or no emission in all \textit{Herschel} far-infrared 
bands (70-500 $\mu$m). This was done to obtain a good baseline for such sensitive observations. Mars was used as a calibrator
to determine the main beam efficiencies of the individual pixels. The
fitted water vapor column was typically around 10 $\mu$m. All the
intensities reported here are on the main beam temperature scale. The
\OI\ and \CII\ observations have a main beam size of
6.3$^{\prime\prime}$ and 14.1$^{\prime\prime}$, respectively, and both
data sets were smoothed to a spectral resolution of 0.3 km
s$^{-1}$. \\
There is no detection of the \CII\ and \OI\ lines (Sec.~\ref{SOFIA})
in the individual pixels (Fig. A.1) or in the array-averaged spectra
(Fig.~\ref{spectraCrestNorth}).  The 3$\sigma$ upper limits for the
two lines are 0.11 K and 0.19 K, equivalent to 7.5$\cdot$10$^{-7}$ erg
s$^{-1}$ cm$^{-2}$ sr$^{-1}$ and 2.1$\cdot$10$^{-5}$ erg s$^{-1}$
cm$^{-2}$ sr$^{-1}$ when assuming a FWHM of 1 km s$^{-1}$, for
\CII\ and \OI, respectively.


\section{Results and analysis} \label{sec:results}

\subsection{The ambient FUV field from a census of nearby stars} \label{sec:fuvstars}

We estimated the FUV field upper limit in Musca using a census of nearby ionising stars \citep[e.g.][]{Schneider2016}. 
For that, we employed the SIMBAD\footnote{http://simbad.u-strasbg.fr/simbad/sim-fsam} database \citep{Wenger2000} and imposed a search for B3 and earlier type stars within a radius of 40$^\circ$ (100 pc), centered on the Musca filament, and a maximal distance of 250 pc. This provided a census of 59 ionising stars including two dominant O9-type stars. Knowing the spectral type of these ionising stars, we estimated the contribution of the FUV emission using model atmospheres\footnote{http://www.oact.inaf.it/castelli/castelli/grids.html} of \citet{Kurucz1979} with solar abundances \citep{Grevesse1998} and log(g) $\sim$4.0 (g is the surface gravity). The temperature of the different spectral types was obtained from \citet{Fitzpatrick2005} and \citet{Pecaut2013}. The distance information and the location on the sky of both Musca and the stars then allows to calculate their relative distances.  
We assumed a r$^{-2}$ decrease of the flux and projected all stars in the plane of the sky, and obtained in this way an ambient FUV field of 3.4 G$_{0}$ \citep{Habing1968} for the Musca cloud as an upper limit. The real field is lower because we did not take into account any extinction, that is, attenuation by diffuse gas and blocking by molecular clumps.

\subsection{Limits on FUV heating from CI \& CII} \label{sec:fuvLimits}

As a second method to constrain the ambient FUV field, we used the
  SOFIA \CII\ upper limit and the APEX \CI\ brightness observed in Musca.
\CII\ is an excellent tracer of the FUV field at low densities as it
is a direct result of the ionising radiation
\citep[e.g.][]{Tielens1985} and its non-detection points towards a low
value for the ambient FUV field.  \CI\ is not a direct product of FUV
ionisation, but PDR models demonstrate that its brightness is also
sensitive to the FUV field
\citep[e.g.][]{Hollenbach1997,Roellig2007}. In this section, we compare the
observed line brightness with predictions from PDR
models using the 2006 version of the PDR toolbox\footnote{http://dustem.astro.umd.edu/pdrt/}
\citep{Kaufman2006,Pound2008}. These plots are expressed in molecular hydrogen density n$_{H_{2}}$, where we assume that all hydrogen atoms are locked in molecular hydrogen.

%
%
%
%
The \CII\ upper limit restricts the FUV field to values $<$ 1 G$_{0}$
for densities n$_{H_{2}}$ $\le$ 10$^{4}$ cm$^{-3}$. This is consistent with the
intensity of the \CI\ line which is so weak that it is below the
minimum value in the PDR toolbox.
%
%
Even with a beam filling value of 30 \% for \CI, which would be low
around the Musca filament, the FUV field remains below 1 G$_{0}$ for typical 
densities of the Musca ambient cloud. Using the Meudon PDR
code\footnote{http://ismdb.obspm.fr/}
\citep{LePetit2006,LeBourlot2012,Bron2014} for the same lines,
confirms that the upper limit on the \CII\ intensity can only be the
result of a FUV field strength $<$ 1 G$_{0}$. With these observational
upper limits for the ambient FUV field, the \citet{Kaufman2006} PDR
models restrict the maximal surface temperature of the cloud to 25
K. This temperature upper limit for gas embedded in such a weak FUV
field is also found in other theoretical models where heating by the
FUV field is taken into account \citep[e.g.][]{Godard2019}, and
consistent with the maximal CO temperature in numerical simulations of
molecular clouds evolving in a FUV field $\le$ 1.7 G$_{0}$
\citep{Glover2016,Clark2019}.

\subsection{CO spectra in Musca}  \label{sec:spectra}
\subsubsection{Three CO velocity components}
Figure \ref{spectraCrestNorth} shows spectra averaged across the northern (top panel) and southern 
(bottom panel) filament for \textit{Herschel} column densities N $>$ 3$\cdot$10$^{21}$ cm$^{-2}$.

The spectra of the CO isotopologues unveil three velocity components that 
show small velocity variations in the north and south. First, there is a single 
component in C$^{18}$O(2-1) at 3.5 km s$^{-1}$ (north) and 3.1 km s$^{-1}$ (south) 
that traces the Musca filament crest \citep{Hacar2016}. In $^{13}$CO(2-1) one 
observes a blueshifted shoulder to the velocity component of the 
filament crest, which is a velocity component related to the strands, namely, 
the dense interface region between the filament crest and the ambient cloud.  
It has a typical velocity of 3.1 km s$^{-1}$ in the north and 2.7 km s$^{-1}$ 
in the south and is not detected in C$^{18}$O with current data. In $^{12}$CO a 
third velocity component is observed, without any clearly detected counterpart in $^{13}$CO or C$^{18}$O, 
which is further blueshifted so that we call it the $'$blueshifted component$'$. The brightness peak of the blueshifted component in the northern map
occurs at 2.8 km s$^{-1}$ and in the southern map at 2.5 km s$^{-1}$. This 
blueshifted component is the focus of this paper.

The individual $^{12}$CO(4-3) spectra in the northern map are
presented in Fig. \ref{singleSpectraCO43}, overlaid on the
\textit{Herschel} column density and temperature map in the top 
panels and together with individual spectra in other CO isotopologues at selected
positions in the bottom panel.  Inspecting the $^{12}$CO(3-2) and
$^{12}$CO(4-3) spectra, we find that these lines are clearly detected
towards the lower column density regions and that this blueshifted
component is present over the entire map, see Fig. \ref{singleSpectraCO43}.

Because our assumption is that the Musca filament is cylindrical, the observed CO emission 
in direction of the crest and strands also contains gas from the ambient cloud along the 
line-of-sight. In particular, the $^{12}$CO(2-1) blueshifted line can have a more significant component 
arising from this gas phase. We can thus not fully exclude effects of high optical depth and self-absorption, so 
we run tests using a two-layer gas model for the CO(2-1) isotopologue lines in order 
to calculate a possible impact of foreground absorption and present the results in Appendix \ref{fittingAppendix}. 
For that, we assumed only 2 CO line components, i.e. the crest component and a shoulder+blueshifted one that it self-absorbed and only apparently shows separate components. 
However, the result of this modelling is that it is not possible to reproduce the observed CO line profiles 
with only two components, so that we are confident in our approach to use 3 separate lines. The best fitting 
model for the blueshifted component of all CO lines in the northern and southern map is the one with the parameters given in Table \ref{FittingTable}. We note from the Table that the fitted $^{12}$CO(2-1) linewidth of the blueshifted component in the south is higher than the linewidth obtained for $^{12}$CO(3-2) and $^{12}$CO(4-3), which have a similar width. It is thus possible that there, the low-J $^{12}$CO(2-1) line traces more material along the line-of-sight from the ambient cloud. 
Inspecting the spectra, see Fig. \ref{spectraCrestNorth}, confirms that $^{12}$CO(3-2) and $^{12}$CO(4-3) only show up at higher velocities and have a smaller linewidth. This difference, and the irregular $^{12}$CO(2-1) shape in the southern map, suggests that $^{12}$CO(2-1) traces the ambient cloud down to lower velocities than the detected $^{12}$CO(4-3) and (3-2) emission. 
We also determined from the averaged $^{12}$CO(2-1) and $^{13}$CO(2-1) spectra that the $^{12}$CO(2-1)/$^{13}$CO(2-1) line ratio in the blueshifted component in the northern and southern map are between 15 and 60, which are expected values for optically thin emission. 
Note that fractionation (see Appendix B) can cause observed values for the $^{12}$CO/$^{13}$CO ratio significantly below 60. This is in agreement with early indications for Musca of significant CO isotopologue abundace variations \citep{Hacar2016} which we further confirm in Paper I.     
  



\begin{table}
\small
\begin{tabular}{c|c|c|c}
line & FWHM (km s$^{-1}$) & $\int$ T dv (K km s$^{-1}$) & v$_{blue}$ (km s$^{-1}$)\\
\hline
\multicolumn{4}{c}{north}\\
\hline
$^{12}$CO(2-1) & 0.35 & 2.0 & 2.7\\
$^{12}$CO(3-2) & 0.31 & 0.98 & 2.8\\
$^{12}$CO(4-3) & 0.42 & 0.73 & 2.9\\
\hline
\multicolumn{4}{c}{south}\\
\hline
$^{12}$CO(2-1) & 0.71 & 4.6 & 2.5\\
$^{12}$CO(3-2) & 0.35 & 1.7 & 2.5\\
$^{12}$CO(4-3) & 0.33 & 0.63 & 2.5\\
\end{tabular}
\caption{Linewidth, integrated brightness and velocity for the average 
$^{12}$CO blueshifted component in the northern and southern map after 
fitting 3 gaussians to the spectrum.}
\label{FittingTable}
\end{table}


\subsubsection{Density for the blueshifted component}
\label{sec: densitySection}

From the sections before, we learned that the blueshifted component is an individual feature 
 towards the crest and strand regions, and that the $^{12}$CO(2-1) line has contributions from the ambient cloud along the line-of-sight. 
 Here, we estimate the typical density and upper limit for this ambient cloud close to the filament. For that we use the density of the fitted Plummer profile at the outer radius of the filament. With the values from \citet{Cox2016}, correcting for a distance of 140 pc for the Musca filament, this gives n$_{H_{2}}$ = 4.0$\cdot$10$^{2}$ cm$^{-3}$ at r = 0.2 pc. 
 An approach to estimate the density upper limit is to combine the maximal column density associated with the ambient cloud, which is N$\sim$10$^{21}$ cm$^{-2}$ \citep{Cox2016}, with the minimal possible size of the ambient cloud. 
 As it is unlikely that the ambient cloud has a smaller size along the line of sight than the filament, this gives a minimal size of 0.4 pc. Using a minimal size of 0.4 pc along the line of sight gives an upper limit of n$_{H_{2}}$ = 8$\cdot$10$^{2}$ cm$^{-3}$ for the ambient cloud.

The \CI\ emission at the velocity of the blueshifted component in Fig. \ref{spectraCrestNorth} is weak, $\sim$1-1.5 K, but it is difficult to constrain this because of the noise. This indicates that \CI\ can trace the ambient cloud down to at least n$_{H_{2}}$ $\sim$ 4$\cdot$10$^{2}$ cm$^{-3}$ in a weak FUV-field. More extensive \CI\ observations will be required to better understand the physical conditions traced by \CI\ in the ISM. This is particularly important for comparison with simulations since the physical conditions traced by \CI\ remain uncertain in simulations \citep[e.g.][]{Glover2015,Franeck2018,Clark2019}.

\subsection{$^{12}$CO(4-3) excess emission in the blueshifted velocity component}
\label{sec: excess}

The $^{12}$CO(3-2) and $^{12}$CO(4-3) transitions have relatively high excitation
temperatures \citep[33 K and 55 K respectively;][]{Mueller2005}. This allows us to investigate the presence of warm gas in the Musca cloud. 
We focus on the blueshifted component for which $^{12}$CO(4-3) is observed over the full map. This could be an indication of quite uniform heating by e.g. a FUV field, though there are indications of spatial variations in the data. However, because of the noise in the spatially resolved map, we will not focus on these possible spatial variations in this paper. To further investigate the heating that leads to the $^{12}$CO(4-3) emission, we study the $^{12}$CO(4-3)/$^{12}$CO(2-1) and $^{12}$CO(4-3)/$^{12}$CO(3-2) brightness temperature ratios. In the previous section, we noted that it is important to take care comparing the integrated brightness of $^{12}$CO(2-1) with $^{12}$CO(3-2) and $^{12}$CO(4-3) in the blueshifted component because of their difference in linewidth. Furthermore, the noise in the $^{12}$CO(3-2) and $^{12}$CO(4-3) spectra makes it impossible to confidently fit a spectrum for every pixel in the map, even when spatially smoothing the data. 
Because of this, we will approach this section by focusing on the peak brightness temperatures. 

\begin{figure}
\includegraphics[width=\hsize]{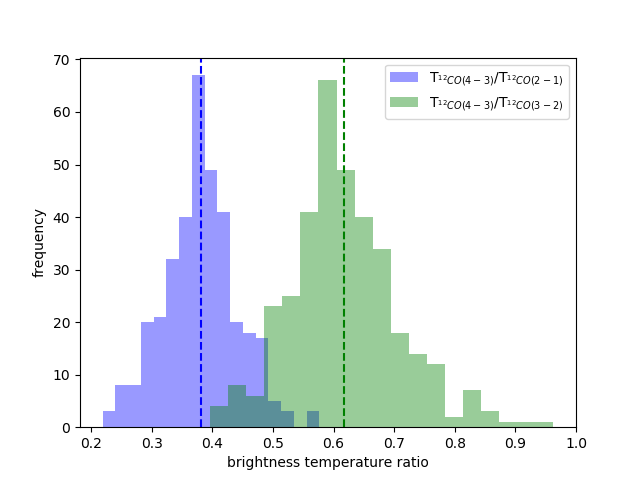}
\caption{Histograms of the $^{12}$CO(4-3)/$^{12}$CO(2-1) (blue) and $^{12}$CO(4-3)/$^{12}$CO(3-2) (green) main beam brightness ratio
  for the pixels in the northern map with a $^{12}$CO(4-3) rms $<$ 0.2
  K. The dashed blue line indicates the average $^{12}$CO(4-3)/$^{12}$CO(2-1) ratio for all pixels,
  and the dashed green line indicates the average $^{12}$CO(4-3)/$^{12}$CO(3-2) ratio.}
\label{rat4321Hist}
\end{figure}

The peak brightness of the blueshifted component occurs near 2.8 km s$^{-1}$ in the northern map and near 2.5 km s$^{-1}$ in the southern map. At these velocities, we determined the brightness temperature ratios for every pixel in the maps. In Fig. \ref{rat4321Hist}, the distribution of the $^{12}$CO(4-3)/$^{12}$CO(2-1) and $^{12}$CO(4-3)/$^{12}$CO(3-2) peak brightness ratios for the northern map are displayed. We determined the average and median values for the various ratios in both maps which are summarized in Tab. \ref{ratioTab}. Overall, there is no large difference between average and median values. For the $^{12}$CO(4-3)/$^{12}$CO(2-1) and $^{12}$CO(4-3)/$^{12}$CO(3-2) ratio in the northern map, we obtain a median value of 0.38 and 0.61 at 2.8 km s$^{-1}$, respectively. In the southern map, the noise of both the $^{12}$CO(4-3) and $^{12}$CO(3-2) transition is higher as it was observed under poorer weather conditions. The median $^{12}$CO(4-3)/$^{12}$CO(2-1) ratio is 0.38 at 2.5 km s$^{-1}$ and the respective value for the $^{12}$CO(4-3)/$^{12}$CO(3-2) ratio is 0.53. 


\begin{table}[]
\tiny
    \centering
    \begin{tabular}{c|c|c|c|c}
        &  \multicolumn{2}{c|}{$^{12}$CO(4-3)/$^{12}$CO(2-1)} & \multicolumn{2}{c}{ $^{12}$CO(4-3)/$^{12}$CO(3-2)}  \\
        & mean & median & mean & median\\
        \hline
        north (2.8 km s$^{-1}$) & 0.38 $\pm$ 0.06 & 0.38 & 0.62 $\pm$ 0.09 &  0.61 \\
         south (2.5 km s$^{-1}$) & 0.39 $\pm$ 0.13 & 0.38 & 0.55 $\pm$ 0.19 &   0.53 \\
    \end{tabular}
    \caption{Mean and median main beam brightness ratios for $^{12}$CO(4-3)/$^{12}$CO(2-1) and $^{12}$CO(4-3)/$^{12}$CO(3-2) in the blueshifted component for the northern and southern map.}
    \label{ratioTab}
\end{table}

\subsubsection{Modeling with the PDR Toolbox}
\begin{figure}
\includegraphics[width=\hsize]{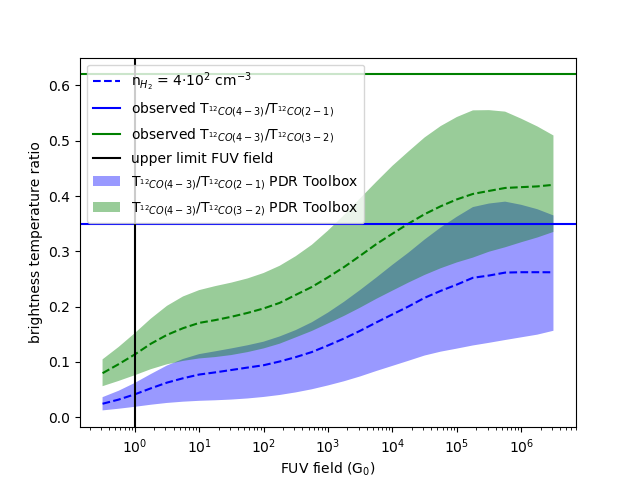}
\caption{Observed $^{12}$CO(4-3)/$^{12}$CO(2-1) and $^{12}$CO(4-3)/$^{12}$CO(3-2) ratios compared with predictions by the PDR Toolbox. The shaded areas, constructed by logarithmic interpolation between the points in the PDR Toolbox density grid, show the evolution of the $^{12}$CO(4-3)/$^{12}$CO(2-1) and $^{12}$CO(4-3)/$^{12}$CO(3-2) brightness temperature   ratio as a function of the FUV field in the
  allowed density range of the Musca cloud. At a typical density n$_{H_{2}}$ = 4$\cdot$10$^{2}$ cm$^{-3}$ of the ambient cloud (dashed line) for the allowed FUV field strength ($<$ 1 G$_{0}$), the predicted line ratios are only a fraction of the observed ones in the blueshifted component.}
\label{ratioPdrToolbox}
\end{figure}

We compared the observed $^{12}$CO(4-3)/$^{12}$CO(2-1) and $^{12}$CO(4-3)/$^{12}$CO(3-2) brightness temperature ratios in the
blueshifted component with predictions using the PDR toolbox in the allowed density range of the ambient cloud in Musca, see Sec. \ref{sec: densitySection}. Figure~\ref{ratioPdrToolbox} shows that for a FUV field strength $<$1 G$_{0}$, the predicted $^{12}$CO(4-3)/$^{12}$CO(2-1) brightness temperature ratio is smaller than 0.05, which is more than a factor 5 lower than observed in Musca. The same is found for the $^{12}$CO(4-3)/$^{12}$CO(3-2) brightness temperature ratio, with predicted ratios of $\sim$ 0.1, which is again more than a factor 5 lower than observed.
It is thus impossible to obtain the observed ratios for the densities and FUV field strength in the ambient Musca cloud that are allowed by \CII, \CI, \OI, and the calculated upper limit from the nearby census of OB stars. 
This strongly suggests that the $^{12}$CO(4-3) emission in the low-column density blueshifted component cannot be explained as a result of FUV heating.


\subsubsection{Modeling with RADEX}

\begin{figure}
\includegraphics[width=\hsize]{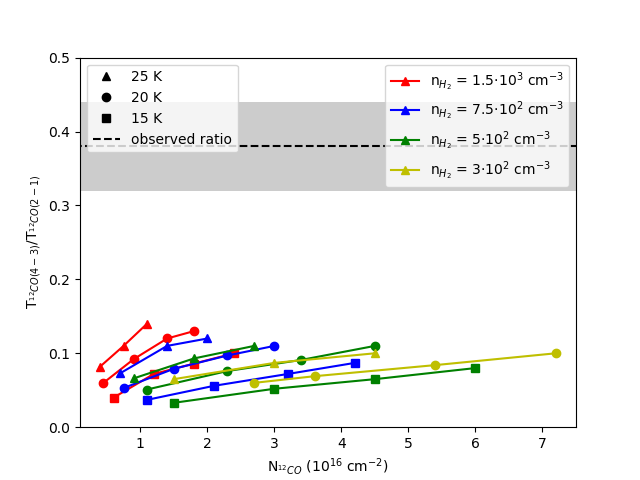}
\caption{$^{12}$CO(4-3)/$^{12}$CO(2-1) brightness temperature ratios predicted with RADEX for the plausible range of physical conditions in the ambient cloud of the Musca filament. The lines connect the RADEX models with identical density and temperature for varying $^{12}$CO column densities. The predicted ratios with RADEX show a huge discrepancy with the observed $^{12}$CO(4-3)/$^{12}$CO(2-1) ratio towards the Musca cloud (indicated by the dashed line). The standard deviation of the observed values from the average is indicated by the grey area. The same is observed for the $^{12}$CO(4-3)/$^{12}$CO(3-2) brightness temperature ratio.}
\label{ratioVsColDens}
\end{figure}

We further investigate whether the $^{12}$CO(4-3)/$^{12}$CO(2-1) and $^{12}$CO(4-3)/$^{12}$CO(3-2) brightness temperature
ratios in the blueshifted component can be reproduced with the non-LTE RADEX code \citep{vanderTak2007}. We use a non-LTE approach because the high critical density (n$_{H_{2}}$ $>$ 5$\cdot$10$^{4}$ cm$^{-3}$) of $^{12}$CO(3-2) and $^{12}$CO(4-3) implies that these lines are subthermally excited. For RADEX analysis we use a FWHM of 0.4 km s$^{-1}$,  temperatures between 15 and 25 K, and
representative densities for the ambient cloud: n$_{H_{2}}$ = 3$\cdot$10$^{2}$, 5$\cdot$10$^{2}$, 7.5$\cdot$10$^{2}$ and 1.5$\cdot$10$^{3}$ cm$^{-3}$.\\ 
For each RADEX model, we also need an upper limit on the $^{12}$CO column density of the blueshifted component. This can be done by calculating with RADEX the $^{13}$CO column density from the $^{13}$CO(2-1) brightness ($\sim$0.6 K, see Fig. \ref{spectraCrestNorth}) for every density and temperature. In a
weak FUV field the [$^{12}$CO]/[$^{13}$CO] abundance ratio is $\le$ 60 \citep[][]{Visser2009,Roellig2013}, which puts an upper limit on the $^{12}$CO column density and opacity. For further analysis, we use RADEX models with a predicted $^{12}$CO(2-1) brightness up to 50\% brighter than observed towards Musca to take into account some uncertainties such as possible opacity broadening of $^{12}$CO(2-1) compared to $^{12}$CO(4-3), a non unity beam filling or different line calibration (uncertainty in $\eta_{\rm mb}$).\\
Studying the $^{12}$CO(4-3)/$^{12}$CO(2-1) and $^{12}$CO(4-3)/$^{12}$CO(3-2) brightness temperature ratios within these limits, we find that the
 ratio increases with increasing density, column density or temperature, see
Fig. \ref{ratioVsColDens}. However, no model is capable to reproduce more than 40\% of the observed $^{12}$CO(4-3)/$^{12}$CO(2-1) and $^{12}$CO(4-3)/$^{12}$CO(3-2) ratios towards the Musca ambient cloud, see Fig. \ref{ratioVsColDens} for the $^{12}$CO(4-3)/$^{12}$CO(2-1) ratio.\\
Taking higher densities for the ambient cloud than put forward in sect. \ref{sec: densitySection}, does not offer a solution for bringing calculations in agreement with observations. Though this can increase the predicted ratio for a fixed temperature and $^{12}$CO column density, this also strongly increases the line brightness temperature. Consequently, one either has to reduce the temperature or $^{12}$CO column density to keep line brightnesses that are not too far off from the observed values. This effort again forces brightness temperature ratios to values found in Fig. \ref{ratioVsColDens}.\\\\
From both approaches, it thus appears
impossible to reproduce the observed $^{12}$CO(4-3)/$^{12}$CO(2-1) and $^{12}$CO(4-3)/$^{12}$CO(3-2)
brightness temperature ratios assuming that there is only collisional and radiative heating.


\begin{figure}
\includegraphics[width=\hsize]{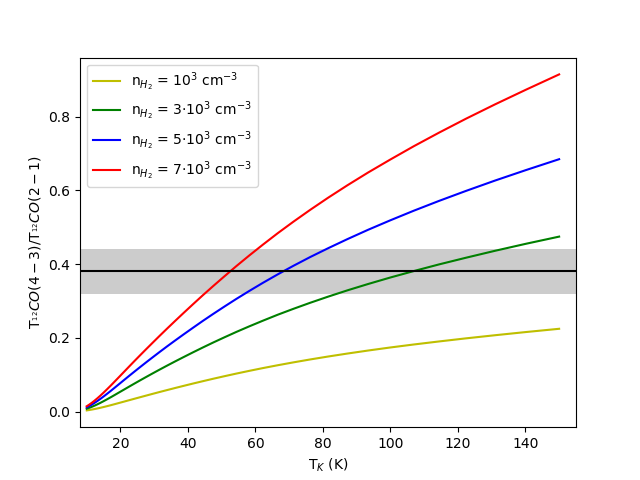}
\caption{Predicted brightness temperature ratio of
  $^{12}$CO(4-3)/$^{12}$CO(2-1) by RADEX as a function of temperature
  for different densities. This demonstrates the need for a warm and dense CO layer to at least reach the observed brightness ratios in the blueshifted component. For each density we plot the ratio for a
  column density such that the predicted brightness of 
  $^{12}$CO(4-3) by RADEX is roughly similar to the observed brightness towards Musca (T$_{mb}$=1.5 - 3 K) at the temperatures that obtain sufficiently high brightness ratios. The lowest three densities (n$_{H_{2}}$ = 10$^{3}$, 3$\cdot$10$^{3}$ and 5$\cdot$10$^{3}$ cm$^{-3}$) use N$_{^{12}{\rm CO}}$ = 1.1$\cdot$10$^{15}$ cm$^{-2}$, and n$_{H_{2}}$ = 7$\cdot$10$^{3}$ cm$^{-3}$ uses N$_{^{12}{\rm CO}}$ = 9$\cdot$10$^{14}$ cm$^{-2}$. The black horizontal line indicates the average observed ratio towards Musca, and the grey area indicates the standard deviation.}
\label{ratioWarmLayer}
\end{figure}

\begin{figure}
\includegraphics[width=\hsize]{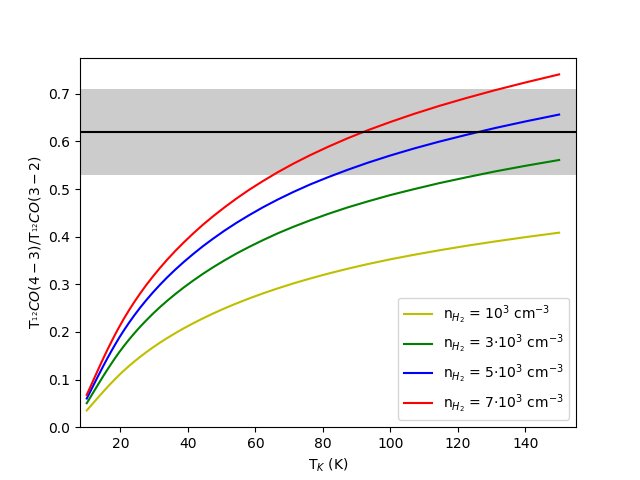}
\caption{Predicted brightness temperature ratio of
  $^{12}$CO(4-3)/$^{12}$CO(3-2) by RADEX for the same models that are presented in Fig. \ref{ratioWarmLayer}. It confirms the need for high temperatures ($>$ 50 K) and densities (5-7$\cdot$10$^{3}$ cm$^{-3}$) to reach the observed excitation of $^{12}$CO(4-3). The black horizontal line indicates the
  average observed ratio towards Musca, and the grey area indicates the standard deviation.}
\label{ratioWarmLayer1}
\end{figure}

\subsection{A warm and dense gas component}
\label{sec: warmGasSection}

\subsubsection{The CO line ratio with RADEX}
In order to increase the $^{12}$CO(4-3)/$^{12}$CO(2-1) and $^{12}$CO(4-3)/$^{12}$CO(3-2) brightness temperature ratio while
keeping the brightness of both lines low enough, one needs to consider a small layer embedded in the diffuse gas that can increase the
$^{12}$CO(4-3) brightness without significantly increasing the $^{12}$CO(2-1) and $^{12}$CO(3-2)
emission. Based on the $^{12}$CO(4-3) excitation conditions, the layer should contain warm gas ($>$ 50 K). 
In fact, this 'layer' can be clumpy, that is, pockets of warm gas that are embedded in more tenuous interclump gas. 
However, modelling such a scenario is not possible with RADEX and out of the scope of this paper. 
Our objective here is to show the existence of a warm gas component. We thus investigated with RADEX the impact
of a warm gas layer by running models with a FWHM of 0.4 km s$^{-1}$ as an overall average, and 
temperatures between 10 and 150 K at different densities. Note that a temperature of 150 K would result in a 
thermal FWHM of 0.5 km s$^{-1}$. This is higher than the observed typical FWHM of 0.4 km s$^{-1}$, but we anticipate that there is also uncertainty on the FWHM so that we work with a kinetic temperature upper limit of 150 K. We additionally restrain ourself to models that manage to reproduce the observed $^{12}$CO(4-3) brightness temperature in Musca\footnote{Using brightness temperatures for modelling 
is less convenient because of the unknown beam filling factors of CO emission  which are eliminated to first order using line ratios. We nevertheless use the $^{12}$CO(4-3) line as an additional - though weak - indicator for the best fitting model.}. Figures \ref{ratioWarmLayer} and \ref{ratioWarmLayer1} show the results for different densities. Both the $^{12}$CO(4-3)/$^{12}$CO(2-1) and $^{12}$CO(4-3)/$^{12}$CO(3-2) ratios indicate that a temperature $>$ 50 K as well as relatively high densities are required to reach the observed ratios. The $^{12}$CO(4-3)/$^{12}$CO(3-2) ratio, in particular, points to high densities and temperature, see Fig. \ref{ratioWarmLayer1}, in order to reach the observed ratios.


 Though the RADEX analysis demonstrates the need for warm and dense gas to obtain the observed $^{12}$CO(4-3) brightness in the blueshifted component, we can not narrow down more precisely the temperature and density range because there are no higher-J CO observations at the moment. This warm and dense gas exactly fits with the predictions for
gas heated by low-velocity shocks \citep[][]{Pon2012,Lesaffre2013}, and so-called slow-type (with regard to their phase velocity) magnetised
shocks with v$_{s}$ = 1-3 km s$^{-1}$ can easily reach temperatures as high as 100 K \citep{Lehmann2016a}. These slow-type shock
models predict a physical size of the warm gas layer around 10$^{15}$
cm, which fits with the estimated physical sizes for the RADEX models. The estimated size, that is the layer thickness, of the models that manage to reach the observed brightness temperature ratios in Figs. \ref{ratioWarmLayer} and \ref{ratioWarmLayer1} are 1.3$\cdot$10$^{15}$ cm and 2.2$\cdot$10$^{15}$ cm, respectively at
n$_{H_{2}}$ = 7$\cdot$10$^{3}$ cm$^{-3}$ (with N$_{\rm ^{12}CO}$ = 9$\cdot$10$^{14}$ cm$^{-2}$) and 5$\cdot$10$^{3}$ cm$^{-3}$ (with N$_{\rm ^{12}CO}$ = 1.1$\cdot$10$^{15}$ cm$^{-2}$).
To calculate the physical size of the RADEX models an abundance of $\big[$H$_{2}\big]$/$\big[^{12}$CO$\big]$ = 10$^{4}$ was used, which is a typical value for weakly irradiated molecular gas.\\
Lastly, we note that this warm gas is observed in both maps and thus
likely universally present around the Musca filament. 

\subsubsection{CI column density in the blueshifted component}
We noted in Fig. \ref{spectraCrestNorth} the presence of \CI\ emission in the blueshifted component with a brightness temperature of the order of 1-1.5 K. From this weak \CI\ emission we here estimate the \CI\ column density. We use a FWHM = 0.4 km s$^{-1}$, a temperature of 20 K, and typical densities n$_{H_{2}}$ = 5-7.5$\cdot$10$^{2}$ cm$^{-3}$ for the ambient Musca cloud with RADEX. This points to N$_{\rm CI}$ $\sim$ 10$^{16}$ cm$^{-2}$, see Tab. \ref{CItable}. Comparing this with the same models for $^{12}$CO emission from the ambient cloud in section \ref{sec: excess}, we find that at least 20\% and possibly up to 50\% of carbon is still found in its atomic form. 

On the other hand, it is observed from Tab. \ref{CItable} that the warm and dense gas layer, necessary to explain the bright $^{12}$CO(4-3) emission, provides a negligible contribution to the \CI\ emission.\\
We thus emphasise again that the blueshifted component has two contributions:\\
- The ambient cloud which gives rise to \CI\ and low-J CO emission.\\
- A warm gas layer (or pockets of warm gas) with little \CI\ emission that is responsible for the bright CO(4-3) line.
\begin{table}
\begin{center}
\begin{tabular}{c|c|c|c}
T (K) & n$_{H_{2}}$ (cm$^{-3}$) & N$_{\rm CI}$ (cm$^{-2}$) & T$_{\rm mb}$ (K)\\
 (1)  & (2)                     & (3)          & (4)    \\
\hline
\multicolumn{4}{c}{ambient cloud}\\
\hline
20 & 5$\cdot$10$^{2}$ & 10$^{16}$ & 1.0\\
20 & 7.5$\cdot$10$^{2}$ & 10$^{16}$ & 1.2 \\
\hline
\multicolumn{4}{c}{warm gas layer}\\
\hline
60 & 7$\cdot$10$^{3}$ & 9$\cdot$10$^{14}$ & 0.16\\
90 & 7$\cdot$10$^{3}$ & 9$\cdot$10$^{14}$ & 0.15\\
\end{tabular}
\caption{Predicted \CI\ brightness temperatures (4), calculated with RADEX for various temperatures 
(1) and densities (2) for the blueshifted velocity component. This indicates a \CI\ column density (3) 
  of $\sim$10$^{16}$ cm$^{-2}$ for the ambient cloud, while the contribution from the warm gas layer to the observed \CI\ emission is
  negligible.}
\label{CItable}
\end{center}
\end{table}

\subsection{Shock models} 

\begin{figure}
\includegraphics[width=\hsize]{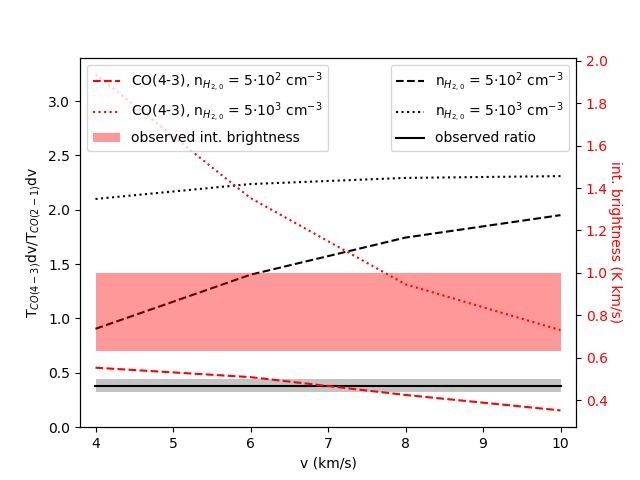}
\caption{$^{12}$CO(4-3)/$^{12}$CO(2-1) integrated brightness ratio (black) for J-type shock models from the Paris-Durham code. The ratio at a shock velocity of 4 km s$^{-1}$ is higher than observed in Musca. This is expected because the ratio decreases towards lower velocities and the $^{12}$CO(2-1) emission in Musca also comes from non shocked gas, such that the observed ratio is a lower limit for the shock models. In red, we show the integrated brightness of $^{12}$CO(4-3) for the same shock models, together with the $^{12}$CO(4-3) integrated brightness interval of the blueshifted component in Musca. The observed values in Musca show a relatively good agreement with the predicted $^{12}$CO(4-3) integrated brightness for a pre-shock density of n$_{H_{2,0}}$ $\sim$ 5$\cdot$10$^{2}$ cm$^{-3}$.}
\label{COparisdurham}
\end{figure}

For an in-depth comparison with shock models, it would be preferable to
have additional observations of mid-to high-J (J$_{up}>$ 4) CO lines
\citep[e.g.][]{Pon2012,Lehmann2016a} since $^{12}$CO(2-1), and possibly some $^{12}$CO(3-2), emission in the blueshifted component also
comes from non-shocked gas (as it is found in synthetic observations
of simulations; see Fig. \ref{simSeamus2}).\\ 
  Because observational studies so far
  focussed on the higher-J CO lines observed with \textit{Herschel} \citep{Pon2014,Larson2015}, it is not well investigated to which extent mid-J CO lines, in particular the $^{12}$CO(4-3) transition,  contribute as a cooling line for shocks. We thus compared our observations to results of the 
  Paris-Durham shock code \citep{Flower2003,Lesaffre2013,Godard2019} that are available 
  for both non-irradiated C- and J-type shocks. \\
Looking into
computed model grids, we find that non-irradiated C-type shocks with a pre-shock density n$_{H_{2,0}}$ $\sim$ 5$\cdot$10$^{2}$ cm$^{-3}$ do not manage to reproduce the observations. Both the integrated intensity of $^{12}$CO(4-3) and the predicted ratios are below the observed values in Musca. However, J-type shocks, which are good first order models for slow-type shocks \citep{Lehmann2016a}, with a pre-shock density n$_{H_{2,0}}$ $\sim$ 5$\cdot$10$^{2}$ cm$^{-3}$ fit with the $^{12}$CO observations, see Fig. \ref{COparisdurham}. The predictions by these models for the \CII, \OI, and \CI\ brightnesses are also in agreement with the observations of Musca (see Appendix \ref{SOFIA}). J-type shocks as a proxy for slow-type
low-velocity shocks are justified since the dynamics of these slow-type shocks are driven by the gas pressure. Note that we used published shock models from the Paris-Durham code for this first comparison which have shock velocities of 4 km s$^{-1}$ and higher, while in Musca the
shock velocity can be lower. A comparison with shock models at lower velocities is work in progress, and will be addressed in a forthcoming paper. \\
The nondimensional parameter b = (B/1$\mu$G)/ $\sqrt{n \text{ cm}^{-3}}$ = 0.1 in the
shock models, with B the component of the magnetic field that is
perpendicular to the direction of shock propogation, is used to model
J-type shocks, and b = 1 is used to model C-type shocks
\citep{Lesaffre2013}.


To perform a first estimate of the shock velocity, we have to take into account that there is an orientation angle for the
shock propagation with respect to the plane of the sky (POS). We will assume shock propagation along
the magnetic field. In this scenario, one can estimate the true shock
velocity when the angle between the magnetic field and the POS is
known, using

\begin{equation}
{\rm v}_{shock} = {\rm v}_{los}/\sin(\gamma),
\end{equation}

where $\gamma$ is the angle between the magnetic field and the POS,
and v$_{los}$ is the observed shock velocity along the line of sight
for which we take 0.4 km s$^{-1}$. In \citet{PlanckArzoumanian2016}, a
value around 25$^{\circ}$ is put forward for the angle between the
magnetic field and the POS. This results in a shock velocity around
0.9 km s$^{-1}$. This fits nicely with the proposed presence of
slow-type low-velocity shocks, which need to have velocities between
the sound speed (c$_{s}$) and v$_{A}$cos($\theta$). With v$_{A}$ the
alfv\'en speed and $\theta$ the angle between the magnetic field and the
propagation direction of the shock \citep{Lehmann2016a}. The expected
alfv\'en speed towards the ambient cloud in Musca is of the order of 2.2
km s$^{-1}$, assuming a 33$\mu$G magnetic field strength at n$_{H_{2}}$ $\sim$ 4$\cdot$10$^{2}$ cm$^{-3}$ \citep{Crutcher2012}. We thus find that
the estimated shock velocity fits with the conditions that allow a
slow-type shock. An angle $<$ 10$^{\circ}$ between the magnetic field and
the plane of the sky is required for a shock velocity $>$ 2.2 km
s$^{-1}$.

Summarising, the CO brightness temperature ratios that we observe
  in Musca are consistent with the predictions from low-velocity shock
  models \citep[e.g.][]{Pon2012,Lehmann2016a}. Low- to mid-J CO lines
  may thus indeed be used to detect low-velocity shocks but more
  observational studies are required to better understand the
  quantitative contribution from shock excitation.\\\\

%
%
Since SiO transitions can be used as a tracer of certain shocks \citep[e.g.][]{Schilke1997,Gusdorf2008a,Gusdorf2008b}, it is worth noting the 
SiO(5-4) line was not detected towards the Musca filament. With RADEX this allows to put a 3$\sigma$ upper limit on the SiO column density of $\sim$ 10$^{14}$ cm$^{-2}$ at the filament crest, assuming n$_{H_{2}}$ = 10$^{4}$ cm$^{-3}$. This implies that less than 0.06 \% of Si is in the form of SiO at the Musca filament crest (see App. \ref{SiOapp} for more information). Taking that all the gas-phase silicon is found in the form of SiO at the crest \citep{Louvet2016}, suggests that all Si is locked in the grains of the
Musca cloud at an early stage of evolution. To explain the presence of SiO in the gas phase, as is found in regions like W43
\citep{Louvet2016}, implies that there is the need for an important dynamical or radiative history in such clouds to release Si from the grains in the
gas phase.

\section{A filament accretion shock signature in simulations}

\begin{figure*}
\begin{center}
\includegraphics[width=0.39\hsize]{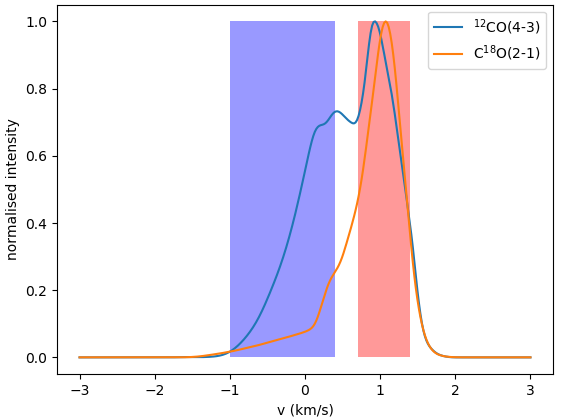}
\includegraphics[width=0.433\hsize]{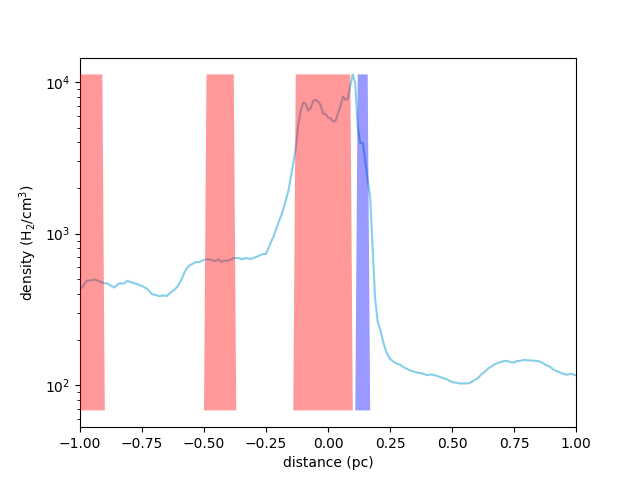}
\includegraphics[width=0.45\hsize]{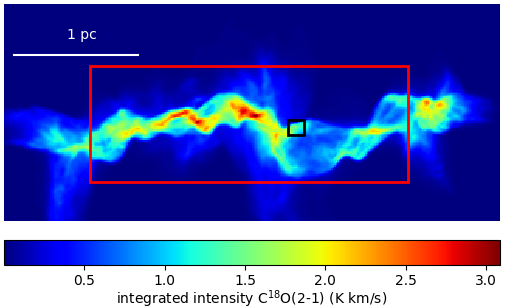}
\end{center}
\caption{\textbf{Bottom}: C$^{18}$O(2-1) integrated intensity map
  of the simulation \citep{Clarke2018}, with the black square indicating the region that is
  studied in this figure. The red box indicates the region displayed in Fig. \ref{tempSkyPlane}. \textbf{Left}: $^{12}$CO(4-3) and C$^{18}$O(2-1)
  spectra extracted from the square which show a blueshifted
  $^{12}$CO(4-3) component (blue) in addition to the component related to the bulk emission of the 
  filament crest (red). The two velocity intervals used for the plot
  on the right are indicated in blue and red. \textbf{Right}: Physical location along the line of sight of the gas in the velocity
  intervals, showing that the larger blueshifted velocity interval
  covers a small region compared to the filament and that it is
  located at the border of this filament where the density strongly
  increases because of a filament accretion shock.}
\label{simSeamus1}
\end{figure*}

\begin{figure}
\includegraphics[width=\hsize]{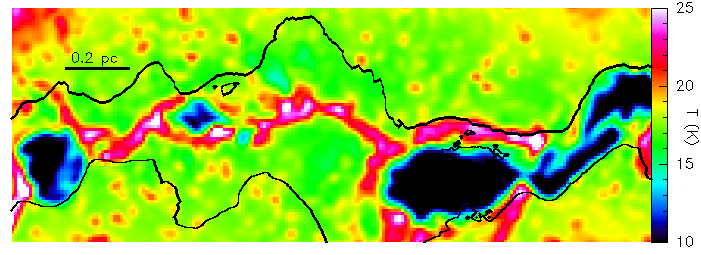}
\caption{Smoothed ($\sigma$ = 1.3 pix) cut at z = 0.05 pc through the temperature cube of the
  simulation, with the black contours (N = 5$\cdot$10$^{21}$ cm$^{-2}$)
  roughly indicating the high column density filament. Note the local
  increase in temperature at the sub-filament border due to the
  accretion shocks and the temperature drop in the sub-filaments after
  the shock compared to the inflowing mass reservoir. Because of smoothing, the temperature peaks of the shocked gas in the map have decreased. The area of the simulation that is covered by this figure is indicated in Fig. \ref{simSeamus1}}
\label{tempSkyPlane}
\end{figure}
\begin{figure*}
\begin{center}
\includegraphics[width=0.45\hsize]{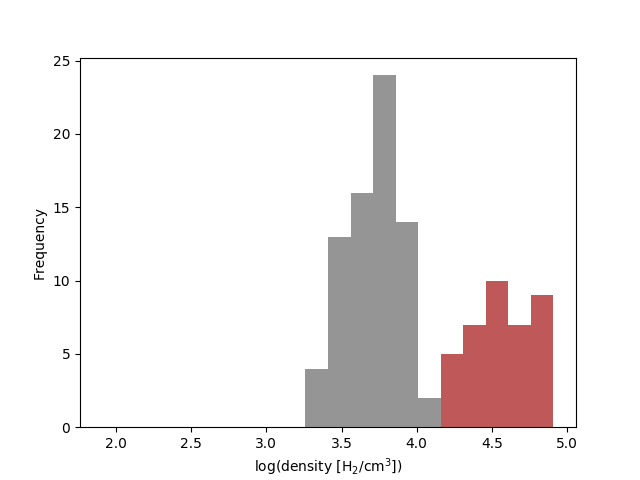}
\includegraphics[width=0.45\hsize]{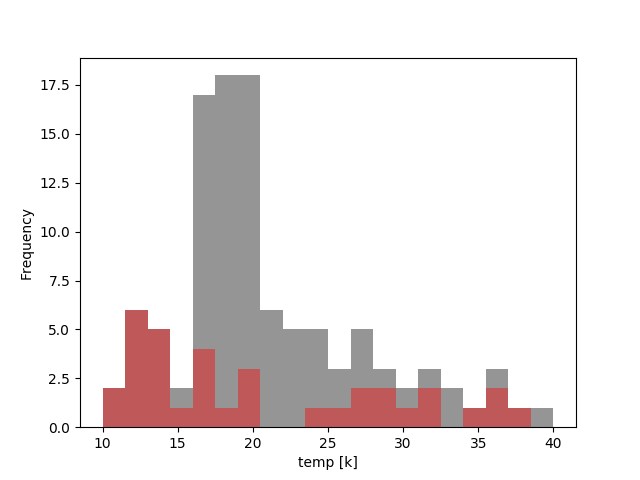}
\end{center}
\caption{\textbf{Left}: Density distribution of the gas with a sufficient CO abundance to be observable in a $^{12}$CO(4-3) excess component. CO is considered detectable if more than half of the carbon is locked in CO. Note that the density distribution is relatively bimodal, with the densest already shocked gas highlighted in red. \textbf{Right}: Temperature distribution of all the CO gas in grey, with in red the temperature of the high density gas highlighted on the left. We find for the dense gas that almost all gas at T = 15 - 25 K has disappeared, leaving only the cold ($<$ 15 K) and warm gas ($>$ 25 K). This demonstrates that post-shock gas is either warm or cooled behind the shocked gas layer compared to the temperature of the inflowing mass reservoir.}
\label{simSeamus2}
\end{figure*}

From the analysis of the observations, we found the presence a of warm gas layer around the dense gas 
(filament crest and strands) in the Musca cloud. These findings are here compared with synthetic observations from a 
converging flow simulation of filament formation \citep{Clarke2018}. We emphasize that this is a qualitative 
comparison to investigate the role of low-velocity shocks in filament formation, because the simulation was set up for simulating slightly denser features such as the filaments in Taurus. Nevertheless, 
we find in the simulations some generic features related to filament accretion that are observed in Musca. In these simulations a radially convergent flow
forms dense filaments with n$_{H_{2}}$ $\gtrsim$ 10$^{4}$ cm$^{-3}$, see Fig. \ref{simSeamus1} and Fig. \ref{simSeamus2}. The filamentary structure, 
that continuously accretes mass from this converging flow, has a size of 3-4 pc. Turbulence in the flow leads to structure formation in the converging flow, causing
inhomogeneities in the accretion and subsequent substructure in the
filament. The simulation includes hydrodynamics, self-gravity, and
heating and cooling coupled with non-equilibrium chemistry. From this
self-consistent CO formation in the simulations, synthetic observations are produced.\\
The filament edge is defined by filament accretion shocks, see for example Fig. \ref{simSeamus1}, which is a low-velocity shock due to mass accretion on an interstellar filament from a converging flow in a molecular cloud. 
For more information on these simulations and
synthetic observations, we refer to \citet{Clarke2017,Clarke2018}.  

No magnetic field is included in the
simulations, which could have an impact on the shock properties
\citep[e.g.][]{Draine1983,Lesaffre2013}. For a shock propagating perpendicular to the magnetic field, this makes the shock wider and decreases the peak temperature. However, slow-type shocks, which can fit the observed emission, tend to
resemble non-magnetised shocks because the dynamics is driven by gas pressure \citep{Lehmann2016a}. In the
simulations the warm post-shock layer is not resolved, as post-shock cooling of a low-velocity J-type shock occurs over small distances \citep[e.g.][]{Lesaffre2013,Whitworth2018}, implying that the gas heated by the shocks is smeared out. Consequently, the temperature of the heated gas is underestimated. Nonetheless, it is generally observed in the simulations that the 3D dense filamentary
structure is confined by a gas layer that shows an increase in temperature as a result of filament accretion shocks.  Figure
\ref{tempSkyPlane} shows a temperature map, cut out of the simulation, that illustrates this temperature increase at the border of the filament, followed by a temperature drop into the filament. 

The good correspondence with synthetic spectra derived from the simulation is remarkable, see Fig. \ref{simSeamus1}. 
We observe the bulk emission of the filament with the C$^{18}$O(2-1) line and, more importantly, a bright blueshifted
component in $^{12}$CO(4-3) with no C$^{18}$O emission that 
connects the velocity of the inflowing cloud and the dense gas. This is similar to what is found in the Musca
filament. In the simulations, this $^{12}$CO(4-3) component corresponds to a strong velocity transition and density increase from a filament accretion shock of the inflowing mass reservoir, see Fig. \ref{simSeamus1}. This accretion shock is responsible for the observed blueshifted $^{12}$CO(4-3) component. It can be wondered how the slightly lower temperatures in the simulation compare to the estimated values for Musca give rise to the same signature. But as mentioned before, the simulations were 
designed for somewhat higher densities than Musca, which increases the $^{12}$CO(4-3) emission and thus compensates for the slightly lower temperatures.


Since the dense filaments in the simulation are confined by an accretion
shock, it is not surprising that the $^{12}$CO(4-3) excess is found
at both observed locations in Musca as it likely confines the entire filament.
Figure \ref{tempSkyPlane} displays 
that behind the heated gas layer by the filament accretion shock there is also a sharp
decrease in temperature from $\sim$15 K to $\sim$10 K. This creation of
cold and dense gas through a filament accretion shock can also be observed in
the temperature histograms of the simulations where the dense gas, 
created by an accretion shock, is either warmer ($>$25 K) or
significantly colder ($\sim$ 10 K) than the inflowing cloud ($\sim$
15-17 K), see Fig. \ref{simSeamus2}. This demonstrates that accretion
shocks from a continuously inflowing cloud, as observed in Musca,
play an essential role in the cooling of the dense star forming
ISM. This cooling is related to the increased shielding of the dense gas and the increased 
cooling rate by CO due to the higher density, since CO is an important coolant of molecular 
clouds \citep[e.g.][]{Goldsmith1978,Whitworth2018}.

\section{Filament formation due to the dissipation of MHD flows}

We have put forward that the $^{12}$CO(4-3)/$^{12}$CO(2-1) and $^{12}$CO(4-3)/$^{12}$CO(3-2) excess in
the blueshifted component can be explained by the presence of warm, dense gas  
that has properties predicted by low-velocity shock models. In
particular the observed emission can fit with slow-type shocks, while simulations show that these slow-type shocks typically occur at relatively low velocities of v$_{\text{s}}$ < 3 km
s$^{-1}$ found in Musca \citep{Lehmann2016b,Park2019}. Furthermore, as the flow near the filament is expected to be well aligned with the magnetic field \citep{Cox2016}, the shock velocity upper limit for slow-type shocks (v$_{A}\cos{\theta}$) reaches a maximum. This increases the possibility for slow-type shocks to occur. A small, but significant,
change in orientation of the magnetic field takes place from the mass
reservoir to the filament in Musca \citep{PlanckArzoumanian2016} which
could be the result of low-velocity shocks \citep{Fogerty2017}. Lastly, the
observed warm and dense gas layer connects the velocity of the
ambient cloud and dense gas in Musca, which indicates
that the accretion shock dissipates the kinetic energy of the supersonic converging flow and increases the cooling rate to form a (trans-)sonic dense and cold filament.\\
If the Musca filament is indeed confined by the accretion shocks, we can do a first check whether the current results give a reasonable estimated mass accretion rate. The mass accretion rate through this low-velocity shock is estimated in Appendix \ref{appMassAccretionRate}, giving a mass accretion rate $\gtrsim$ 20 M$_{\odot}$ pc$^{-1}$ Myr$^{-1}$. This is similar to the typical values reported for mass inflow towards filaments that will form low-mass stars \citep[e.g.][]{Palmeirim2013}, and would imply that Musca can accrete its current mass in 1 Myr.

\section{Summary}

In this paper we present observations of \CII\ and \OI\ with SOFIA, and observations
of $^{12}$CO(4-3), $^{12}$CO(3-2) and \CI\ with APEX towards the Musca filament. 
Studying these important cooling lines, we find that Musca has an extremely weak FUV field ($<$ 1 G$_{0}$), confirmed by 
determining the FUV field with a census of ionising stars. We further estimated a density n$_{H_{2}}$ $\sim$ 5$\cdot$10$^{2}$ cm$^{-3}$ for the ambient gas near the Musca filament, and found that 20 to 50\% of carbon in the ambient cloud is still in atomic form. In the interface region between the ambient cloud and dense gas in
the crest and strands, blueshifted excess emission was found with the $^{12}$CO(4-3), (3-2), and (2-1) lines. \\ 
A non-LTE radiative transfer study with RADEX points to the existence of a small warm ($\gtrsim$50 K) gas layer at relatively high density (n$_{H_{2}} \ge$ 3$\cdot$10$^{3}$ cm$^{-3}$). This $'$layer$'$ can actually be clumpy with pockets of warm gas embedded in a cooler interclump phase. This excess emission fits with predictions for non-irradiated slow-type low-velocity shock models with a pre-shock density n$_{H_{2,0}}\sim5\cdot$10$^{2}$ cm$^{-3}$. The estimated shock velocity also fits with the shock velocity interval for slow-type shocks.\\
Comparing the location of the shock emission in the spectra with synthetic observations in simulations of filament formation, we obtain
a scenario where the dense Musca filament is formed by low-velocity filament accretion shocks in a colliding flow.
These low-velocity MHD filament accretion shocks dissipate the supersonic kinetic energy of the colliding flow, creating a (trans-)sonic dense filament. Because of increased shielding and efficient CO cooling in the dense post-shock gas, these accretion shocks play an important role in
cooling the dense filamentary ISM that can form stars in the near future.

\begin{acknowledgements}
      We thank the anonymous referee for a critical review of the paper and providing useful comments that improved the quality and clarity of this paper. This work was supported by the Agence National de Recherche (ANR/France) and the Deutsche Forschungsgemeinschaft (DFG/Germany) through the project GENESIS (ANR-16-CE92-0035-01/DFG1591/2-1). L.B. also acknowledges support from the Région Nouvelle-Aquitaine. N.S. acknowledges support from the BMBF, Projekt Number 50OR1714 (MOBS-MOdellierung von Beobachtungsdaten SOFIA). We also appreciate support by the German Deutsche Forschungsgemeinschaft, DFG project number SFB 956. We thank M. R\"ollig, V. Ossenkopf-Okada and J. Stutzki for fruitful discussions on PDR and shock excitation, and self-absorption effects. We also thank F. Wyrowski for providing information to correct for the frequency shift in the APEX FLASH observations. This work is based on observations made with the NASA/DLR Stratospheric Observatory for Infrared Astronomy (SOFIA). SOFIA is jointly operated by the Universities Space Research Association, Inc. (DSI) under DLR contract 50 OK 0901 to the University of Stuttgart. This publication is based on data acquired with the Atacama Pathfinder Experiment (APEX) under programme IDs 0100.C-0825(A), 0101.F-9511(A) and 0102.F-9503(A). APEX is a collaboration between the Max-Planck-Institut fur Radioastronomie, the European Southern Observatory, and the Onsala Space Observatory. This research has made use of the SIMBAD database, operated at CDS, Strasbourg, France.
\end{acknowledgements}

%
   \bibliographystyle{aa} 
   \bibliography{template.bib} 

\begin{thebibliography}{99}
\expandafter\ifx\csname natexlab\endcsname\relax\def\natexlab#1{#1}\fi

\bibitem[{{Alves de Oliveira} {et~al.}(2014){Alves de Oliveira}, {Schneider},
  {Mer{\'\i}n}, {Prusti}, {Ribas}, {Cox}, {Vavrek}, {K{\"o}nyves},
  {Arzoumanian}, {Puga}, {Pilbratt}, {K{\'o}sp{\'a}l}, {Andr{\'e}}, {Didelon},
  {Men'shchikov}, {Royer}, {Waelkens}, {Bontemps}, {Winston}, \&
  {Spezzi}}]{AlvesDeOliveira2014}
{Alves de Oliveira}, C., {Schneider}, N., {Mer{\'\i}n}, B., {et~al.} 2014,
  \aap, 568, A98

\bibitem[{{Andr{\'e}} {et~al.}(2014){Andr{\'e}}, {Di Francesco},
  {Ward-Thompson}, {Inutsuka}, {Pudritz}, \& {Pineda}}]{Andre2014}
{Andr{\'e}}, P., {Di Francesco}, J., {Ward-Thompson}, D., {et~al.} 2014, in
  Protostars and Planets VI, 27

\bibitem[{{Andr{\'e}} {et~al.}(2010){Andr{\'e}}, {Men'shchikov}, {Bontemps},
  {K{\"o}nyves}, {Motte}, {Schneider}, {Didelon}, {Minier}, {Saraceno},
  {Ward-Thompson}, {di Francesco}, {White}, {Molinari}, {Testi}, {Abergel},
  {Griffin}, {Henning}, {Royer}, {Mer{\'{\i}}n}, {Vavrek}, {Attard},
  {Arzoumanian}, {Wilson}, {Ade}, {Aussel}, {Baluteau}, {Benedettini},
  {Bernard}, {Blommaert}, {Cambr{\'e}sy}, {Cox}, {di Giorgio}, {Hargrave},
  {Hennemann}, {Huang}, {Kirk}, {Krause}, {Launhardt}, {Leeks}, {Le Pennec},
  {Li}, {Martin}, {Maury}, {Olofsson}, {Omont}, {Peretto}, {Pezzuto}, {Prusti},
  {Roussel}, {Russeil}, {Sauvage}, {Sibthorpe}, {Sicilia-Aguilar}, {Spinoglio},
  {Waelkens}, {Woodcraft}, \& {Zavagno}}]{Andre2010}
{Andr{\'e}}, P., {Men'shchikov}, A., {Bontemps}, S., {et~al.} 2010, \aap, 518,
  L102

\bibitem[{{Arzoumanian} {et~al.}(2011){Arzoumanian}, {Andr{\'e}}, {Didelon},
  {K{\"o}nyves}, {Schneider}, {Men'shchikov}, {Sousbie}, {Zavagno}, {Bontemps},
  {di Francesco}, {Griffin}, {Hennemann}, {Hill}, {Kirk}, {Martin}, {Minier},
  {Molinari}, {Motte}, {Peretto}, {Pezzuto}, {Spinoglio}, {Ward-Thompson},
  {White}, \& {Wilson}}]{Arzoumanian2011}
{Arzoumanian}, D., {Andr{\'e}}, P., {Didelon}, P., {et~al.} 2011, \aap, 529, L6

\bibitem[{{Bisbas} {et~al.}(2017){Bisbas}, {Tanaka}, {Tan}, {Wu}, \&
  {Nakamura}}]{Bisbas2017}
{Bisbas}, T.~G., {Tanaka}, K.~E.~I., {Tan}, J.~C., {Wu}, B., \& {Nakamura}, F.
  2017, \apj, 850, 23

\bibitem[{{Bron} {et~al.}(2014){Bron}, {Le Bourlot}, \& {Le Petit}}]{Bron2014}
{Bron}, E., {Le Bourlot}, J., \& {Le Petit}, F. 2014, \aap, 569, A100

\bibitem[{{Cartledge} {et~al.}(2004){Cartledge}, {Lauroesch}, {Meyer}, \&
  {Sofia}}]{Cartledge2004}
{Cartledge}, S.~I.~B., {Lauroesch}, J.~T., {Meyer}, D.~M., \& {Sofia}, U.~J.
  2004, \apj, 613, 1037

\bibitem[{{Clark} {et~al.}(2019){Clark}, {Glover}, {Ragan}, \&
  {Duarte-Cabral}}]{Clark2019}
{Clark}, P.~C., {Glover}, S. C.~O., {Ragan}, S.~E., \& {Duarte-Cabral}, A.
  2019, \mnras, 486, 4622

\bibitem[{{Clarke} {et~al.}(2017){Clarke}, {Whitworth}, {Duarte-Cabral}, \&
  {Hubber}}]{Clarke2017}
{Clarke}, S.~D., {Whitworth}, A.~P., {Duarte-Cabral}, A., \& {Hubber}, D.~A.
  2017, \mnras, 468, 2489

\bibitem[{{Clarke} {et~al.}(2018){Clarke}, {Whitworth}, {Spowage},
  {Duarte-Cabral}, {Suri}, {Jaffa}, {Walch}, \& {Clark}}]{Clarke2018}
{Clarke}, S.~D., {Whitworth}, A.~P., {Spowage}, R.~L., {et~al.} 2018, \mnras,
  479, 1722

\bibitem[{{Cox} {et~al.}(2016){Cox}, {Arzoumanian}, {Andr{\'e}}, {Rygl},
  {Prusti}, {Men'shchikov}, {Royer}, {K{\'o}sp{\'a}l}, {Palmeirim}, {Ribas},
  {K{\"o}nyves}, {Bernard}, {Schneider}, {Bontemps}, {Merin}, {Vavrek}, {Alves
  de Oliveira}, {Didelon}, {Pilbratt}, \& {Waelkens}}]{Cox2016}
{Cox}, N.~L.~J., {Arzoumanian}, D., {Andr{\'e}}, P., {et~al.} 2016, \aap, 590,
  A110

\bibitem[{{Crutcher}(2012)}]{Crutcher2012}
{Crutcher}, R.~M. 2012, \araa, 50, 29

\bibitem[{{Csengeri} {et~al.}(2016){Csengeri}, {Leurini}, {Wyrowski},
  {Urquhart}, {Menten}, {Walmsley}, {Bontemps}, {Wienen}, {Beuther}, {Motte},
  {Nguyen-Luong}, {Schilke}, {Schuller}, {Zavagno}, \& {Sanna}}]{Csengeri2016}
{Csengeri}, T., {Leurini}, S., {Wyrowski}, F., {et~al.} 2016, \aap, 586, A149

\bibitem[{{Draine} {et~al.}(1983){Draine}, {Roberge}, \&
  {Dalgarno}}]{Draine1983}
{Draine}, B.~T., {Roberge}, W.~G., \& {Dalgarno}, A. 1983, \apj, 264, 485

\bibitem[{{Duarte-Cabral} {et~al.}(2014){Duarte-Cabral}, {Bontemps}, {Motte},
  {Gusdorf}, {Csengeri}, {Schneider}, \& {Louvet}}]{DuarteCabral2014}
{Duarte-Cabral}, A., {Bontemps}, S., {Motte}, F., {et~al.} 2014, \aap, 570, A1

\bibitem[{{Federrath}(2016)}]{Federrath2016}
{Federrath}, C. 2016, \mnras, 457, 375

\bibitem[{{Fitzpatrick} \& {Massa}(2005)}]{Fitzpatrick2005}
{Fitzpatrick}, E.~L. \& {Massa}, D. 2005, \aj, 129, 1642

\bibitem[{{Flower} \& {Pineau des For{\^e}ts}(2003)}]{Flower2003}
{Flower}, D.~R. \& {Pineau des For{\^e}ts}, G. 2003, \mnras, 343, 390

\bibitem[{{Flower} \& {Pineau Des For{\^e}ts}(2010)}]{Flower2010}
{Flower}, D.~R. \& {Pineau Des For{\^e}ts}, G. 2010, \mnras, 406, 1745

\bibitem[{{Fogerty} {et~al.}(2017){Fogerty}, {Carroll-Nellenback}, {Frank},
  {Heitsch}, \& {Pon}}]{Fogerty2017}
{Fogerty}, E., {Carroll-Nellenback}, J., {Frank}, A., {Heitsch}, F., \& {Pon},
  A. 2017, \mnras, 470, 2938

\bibitem[{{Franco}(1991)}]{Franco1991}
{Franco}, G.~A.~P. 1991, \aap, 251, 581

\bibitem[{{Franeck} {et~al.}(2018){Franeck}, {Walch}, {Seifried}, {Clarke},
  {Ossenkopf-Okada}, {Glover}, {Klessen}, {Girichidis}, {Naab}, {W{\"u}nsch},
  {Clark}, {Pellegrini}, \& {Peters}}]{Franeck2018}
{Franeck}, A., {Walch}, S., {Seifried}, D., {et~al.} 2018, \mnras, 481, 4277

\bibitem[{{Glover} {et~al.}(2015){Glover}, {Clark}, {Micic}, \&
  {Molina}}]{Glover2015}
{Glover}, S. C.~O., {Clark}, P.~C., {Micic}, M., \& {Molina}, F. 2015, \mnras,
  448, 1607

\bibitem[{{Glover} \& {Smith}(2016)}]{Glover2016}
{Glover}, S. C.~O. \& {Smith}, R.~J. 2016, \mnras, 462, 3011

\bibitem[{{Godard} {et~al.}(2019){Godard}, {Pineau des For{\^e}ts}, {Lesaffre},
  {Lehmann}, {Gusdorf}, \& {Falgarone}}]{Godard2019}
{Godard}, B., {Pineau des For{\^e}ts}, G., {Lesaffre}, P., {et~al.} 2019, arXiv
  e-prints [\eprint[arXiv]{1901.04273}]

\bibitem[{{Goldsmith}(2001)}]{Goldsmith2001}
{Goldsmith}, P.~F. 2001, \apj, 557, 736

\bibitem[{{Goldsmith} {et~al.}(2008){Goldsmith}, {Heyer}, {Narayanan}, {Snell},
  {Li}, \& {Brunt}}]{Goldsmith2008}
{Goldsmith}, P.~F., {Heyer}, M., {Narayanan}, G., {et~al.} 2008, \apj, 680, 428

\bibitem[{{Goldsmith} \& {Langer}(1978)}]{Goldsmith1978}
{Goldsmith}, P.~F. \& {Langer}, W.~D. 1978, \apj, 222, 881

\bibitem[{{Goldsmith} \& {Langer}(1999)}]{Goldsmith1999}
{Goldsmith}, P.~F. \& {Langer}, W.~D. 1999, \apj, 517, 209

\bibitem[{{G{\'o}mez} \& {V{\'a}zquez-Semadeni}(2014)}]{Gomez2014}
{G{\'o}mez}, G.~C. \& {V{\'a}zquez-Semadeni}, E. 2014, \apj, 791, 124

\bibitem[{{Grevesse} \& {Sauval}(1998)}]{Grevesse1998}
{Grevesse}, N. \& {Sauval}, A.~J. 1998, \ssr, 85, 161

\bibitem[{{Guevara} {et~al.}(2020){Guevara}, {Stutzki}, {Ossenkopf-Okada},
  {Simon}, {P{\'e}rez-Beaupuits}, {Beuther}, {Bihr}, {Higgins}, {Graf}, \&
  {G{\"u}sten}}]{Guevara2020}
{Guevara}, C., {Stutzki}, J., {Ossenkopf-Okada}, V., {et~al.} 2020, \aap, 636,
  A16

\bibitem[{{Gusdorf} {et~al.}(2017){Gusdorf}, {Anderl}, {Lefloch}, {Leurini},
  {Wiesemeyer}, {G{\"u}sten}, {Benedettini}, {Codella}, {Godard},
  {G{\'o}mez-Ruiz}, {Jacobs}, {Kristensen}, {Lesaffre}, {Pineau des
  For{\^e}ts}, \& {Lis}}]{Gusdorf2017}
{Gusdorf}, A., {Anderl}, S., {Lefloch}, B., {et~al.} 2017, \aap, 602, A8

\bibitem[{{Gusdorf} {et~al.}(2008{\natexlab{a}}){Gusdorf}, {Cabrit}, {Flower},
  \& {Pineau Des For{\^e}ts}}]{Gusdorf2008a}
{Gusdorf}, A., {Cabrit}, S., {Flower}, D.~R., \& {Pineau Des For{\^e}ts}, G.
  2008{\natexlab{a}}, \aap, 482, 809

\bibitem[{{Gusdorf} {et~al.}(2008{\natexlab{b}}){Gusdorf}, {Pineau Des
  For{\^e}ts}, {Cabrit}, \& {Flower}}]{Gusdorf2008b}
{Gusdorf}, A., {Pineau Des For{\^e}ts}, G., {Cabrit}, S., \& {Flower}, D.~R.
  2008{\natexlab{b}}, \aap, 490, 695

\bibitem[{{G{\"u}sten} {et~al.}(2006){G{\"u}sten}, {Nyman}, {Schilke},
  {Menten}, {Cesarsky}, \& {Booth}}]{Guesten2006}
{G{\"u}sten}, R., {Nyman}, L.~{\AA}., {Schilke}, P., {et~al.} 2006, \aap, 454,
  L13

\bibitem[{{Habing}(1968)}]{Habing1968}
{Habing}, H.~J. 1968, \bain, 19, 421

\bibitem[{{Hacar} {et~al.}(2016){Hacar}, {Kainulainen}, {Tafalla}, {Beuther},
  \& {Alves}}]{Hacar2016}
{Hacar}, A., {Kainulainen}, J., {Tafalla}, M., {Beuther}, H., \& {Alves}, J.
  2016, \aap, 587, A97

\bibitem[{{Hennebelle}(2013)}]{Hennebelle2013}
{Hennebelle}, P. 2013, \aap, 556, A153

\bibitem[{{Henning} {et~al.}(2010){Henning}, {Linz}, {Krause}, {Ragan},
  {Beuther}, {Launhardt}, {Nielbock}, \& {Vasyunina}}]{Henning2010}
{Henning}, T., {Linz}, H., {Krause}, O., {et~al.} 2010, \aap, 518, L95

\bibitem[{{Heyer} {et~al.}(2016){Heyer}, {Goldsmith}, {Y{\i}ld{\i}z}, {Snell},
  {Falgarone}, \& {Pineda}}]{Heyer2016}
{Heyer}, M., {Goldsmith}, P.~F., {Y{\i}ld{\i}z}, U.~A., {et~al.} 2016, \mnras,
  461, 3918

\bibitem[{{Hincelin} {et~al.}(2011){Hincelin}, {Wakelam}, {Hersant},
  {Guilloteau}, {Loison}, {Honvault}, \& {Troe}}]{Hincelin2011}
{Hincelin}, U., {Wakelam}, V., {Hersant}, F., {et~al.} 2011, \aap, 530, A61

\bibitem[{{Hollenbach} \& {McKee}(1989)}]{Hollenbach1989}
{Hollenbach}, D. \& {McKee}, C.~F. 1989, \apj, 342, 306

\bibitem[{{Hollenbach} \& {Tielens}(1997)}]{Hollenbach1997}
{Hollenbach}, D.~J. \& {Tielens}, A.~G.~G.~M. 1997, \araa, 35, 179

\bibitem[{{Inoue} {et~al.}(2018){Inoue}, {Hennebelle}, {Fukui}, {Matsumoto},
  {Iwasaki}, \& {Inutsuka}}]{Inoue2018}
{Inoue}, T., {Hennebelle}, P., {Fukui}, Y., {et~al.} 2018, \pasj, 70, S53

\bibitem[{{Ivlev} {et~al.}(2019){Ivlev}, {Silsbee}, {Sipil{\"a}}, \&
  {Caselli}}]{Ivlev2019}
{Ivlev}, A.~V., {Silsbee}, K., {Sipil{\"a}}, O., \& {Caselli}, P. 2019, \apj,
  884, 176

\bibitem[{{Jappsen} {et~al.}(2005){Jappsen}, {Klessen}, {Larson}, {Li}, \& {Mac
  Low}}]{Jappsen2005}
{Jappsen}, A.~K., {Klessen}, R.~S., {Larson}, R.~B., {Li}, Y., \& {Mac Low},
  M.~M. 2005, \aap, 435, 611

\bibitem[{{Jenkins}(2009)}]{Jenkins2009}
{Jenkins}, E.~B. 2009, \apj, 700, 1299

\bibitem[{{Jim{\'e}nez-Serra} {et~al.}(2010){Jim{\'e}nez-Serra}, {Caselli},
  {Tan}, {Hernandez}, {Fontani}, {Butler}, \& {van Loo}}]{JimenezSerra2010}
{Jim{\'e}nez-Serra}, I., {Caselli}, P., {Tan}, J.~C., {et~al.} 2010, \mnras,
  406, 187

\bibitem[{{Kainulainen} {et~al.}(2016){Kainulainen}, {Hacar}, {Alves},
  {Beuther}, {Bouy}, \& {Tafalla}}]{Kainulainen2016}
{Kainulainen}, J., {Hacar}, A., {Alves}, J., {et~al.} 2016, \aap, 586, A27

\bibitem[{{Kaufman} {et~al.}(2006){Kaufman}, {Wolfire}, \&
  {Hollenbach}}]{Kaufman2006}
{Kaufman}, M.~J., {Wolfire}, M.~G., \& {Hollenbach}, D.~J. 2006, \apj, 644, 283

\bibitem[{Klein {et~al.}(2014)Klein, Ciechanowicz, Leinz, Heyminck, Gusten,
  Kasemann, Wunsch, Maier, \& Sekimoto}]{Klein2014}
Klein, T., Ciechanowicz, M., Leinz, C., {et~al.} 2014, Terahertz Science and
  Technology, IEEE Transactions on, 4, 588

\bibitem[{{K{\"o}nyves} {et~al.}(2015){K{\"o}nyves}, {Andr{\'e}},
  {Men'shchikov}, {Palmeirim}, {Arzoumanian}, {Schneider}, {Roy}, {Didelon},
  {Maury}, {Shimajiri}, {Di Francesco}, {Bontemps}, {Peretto}, {Benedettini},
  {Bernard}, {Elia}, {Griffin}, {Hill}, {Kirk}, {Ladjelate}, {Marsh}, {Martin},
  {Motte}, {Nguy{\^e}n Luong}, {Pezzuto}, {Roussel}, {Rygl}, {Sadavoy},
  {Schisano}, {Spinoglio}, {Ward-Thompson}, \& {White}}]{Konyves2015}
{K{\"o}nyves}, V., {Andr{\'e}}, P., {Men'shchikov}, A., {et~al.} 2015, \aap,
  584, A91

\bibitem[{{Kurucz}(1979)}]{Kurucz1979}
{Kurucz}, R.~L. 1979, \apjs, 40, 1

\bibitem[{{Larson} {et~al.}(2015){Larson}, {Evans}, {Green}, \&
  {Yang}}]{Larson2015}
{Larson}, R.~L., {Evans}, Neal~J., I., {Green}, J.~D., \& {Yang}, Y.-L. 2015,
  \apj, 806, 70

\bibitem[{{Le Bourlot} {et~al.}(2012){Le Bourlot}, {Le Petit}, {Pinto},
  {Roueff}, \& {Roy}}]{LeBourlot2012}
{Le Bourlot}, J., {Le Petit}, F., {Pinto}, C., {Roueff}, E., \& {Roy}, F. 2012,
  \aap, 541, A76

\bibitem[{{Le Petit} {et~al.}(2006){Le Petit}, {Nehm{\'e}}, {Le Bourlot}, \&
  {Roueff}}]{LePetit2006}
{Le Petit}, F., {Nehm{\'e}}, C., {Le Bourlot}, J., \& {Roueff}, E. 2006, \apjs,
  164, 506

\bibitem[{{Lehmann} {et~al.}(2016){Lehmann}, {Federrath}, \&
  {Wardle}}]{Lehmann2016b}
{Lehmann}, A., {Federrath}, C., \& {Wardle}, M. 2016, \mnras, 463, 1026

\bibitem[{{Lehmann} \& {Wardle}(2016)}]{Lehmann2016a}
{Lehmann}, A. \& {Wardle}, M. 2016, \mnras, 455, 2066

\bibitem[{{Lesaffre} {et~al.}(2013){Lesaffre}, {Pineau des For{\^e}ts},
  {Godard}, {Guillard}, {Boulanger}, \& {Falgarone}}]{Lesaffre2013}
{Lesaffre}, P., {Pineau des For{\^e}ts}, G., {Godard}, B., {et~al.} 2013, \aap,
  550, A106

\bibitem[{{Liszt}(2017)}]{Liszt2017}
{Liszt}, H.~S. 2017, \apj, 835, 138

\bibitem[{{Liszt} {et~al.}(2006){Liszt}, {Lucas}, \& {Pety}}]{Liszt2006}
{Liszt}, H.~S., {Lucas}, R., \& {Pety}, J. 2006, \aap, 448, 253

\bibitem[{{Liszt} \& {Ziurys}(2012)}]{Liszt2012}
{Liszt}, H.~S. \& {Ziurys}, L.~M. 2012, \apj, 747, 55

\bibitem[{{Louvet} {et~al.}(2016){Louvet}, {Motte}, {Gusdorf}, {Nguy{\^e}n
  Luong}, {Lesaffre}, {Duarte-Cabral}, {Maury}, {Schneider}, {Hill}, {Schilke},
  \& {Gueth}}]{Louvet2016}
{Louvet}, F., {Motte}, F., {Gusdorf}, A., {et~al.} 2016, \aap, 595, A122

\bibitem[{{Malinen} {et~al.}(2016){Malinen}, {Montier}, {Montillaud}, {Juvela},
  {Ristorcelli}, {Clark}, {Bern{\'e}}, {Bernard}, {Pelkonen}, \&
  {Collins}}]{Malinen2016}
{Malinen}, J., {Montier}, L., {Montillaud}, J., {et~al.} 2016, \mnras, 460,
  1934

\bibitem[{{Marsh} {et~al.}(2016){Marsh}, {Kirk}, {Andr{\'e}}, {Griffin},
  {K{\"o}nyves}, {Palmeirim}, {Men'shchikov}, {Ward-Thompson}, {Benedettini},
  {Bresnahan}, {di Francesco}, {Elia}, {Motte}, {Peretto}, {Pezzuto}, {Roy},
  {Sadavoy}, {Schneider}, {Spinoglio}, \& {White}}]{Marsh2016}
{Marsh}, K.~A., {Kirk}, J.~M., {Andr{\'e}}, P., {et~al.} 2016, \mnras, 459, 342

\bibitem[{{Meyer} {et~al.}(1998){Meyer}, {Jura}, \& {Cardelli}}]{Meyer1998}
{Meyer}, D.~M., {Jura}, M., \& {Cardelli}, J.~A. 1998, \apj, 493, 222

\bibitem[{{Molinari} {et~al.}(2010){Molinari}, {Swinyard}, {Bally}, {Barlow},
  {Bernard}, {Martin}, {Moore}, {Noriega-Crespo}, {Plume}, {Testi}, {Zavagno},
  {Abergel}, {Ali}, {Anderson}, {Andr{\'e}}, {Baluteau}, {Battersby},
  {Beltr{\'a}n}, {Benedettini}, {Billot}, {Blommaert}, {Bontemps}, {Boulanger},
  {Brand}, {Brunt}, {Burton}, {Calzoletti}, {Carey}, {Caselli}, {Cesaroni},
  {Cernicharo}, {Chakrabarti}, {Chrysostomou}, {Cohen}, {Compiegne}, {de
  Bernardis}, {de Gasperis}, {di Giorgio}, {Elia}, {Faustini}, {Flagey},
  {Fukui}, {Fuller}, {Ganga}, {Garcia-Lario}, {Glenn}, {Goldsmith}, {Griffin},
  {Hoare}, {Huang}, {Ikhenaode}, {Joblin}, {Joncas}, {Juvela}, {Kirk},
  {Lagache}, {Li}, {Lim}, {Lord}, {Marengo}, {Marshall}, {Masi}, {Massi},
  {Matsuura}, {Minier}, {Miville-Desch{\^e}nes}, {Montier}, {Morgan}, {Motte},
  {Mottram}, {M{\"u}ller}, {Natoli}, {Neves}, {Olmi}, {Paladini}, {Paradis},
  {Parsons}, {Peretto}, {Pestalozzi}, {Pezzuto}, {Piacentini}, {Piazzo},
  {Polychroni}, {Pomar{\`e}s}, {Popescu}, {Reach}, {Ristorcelli}, {Robitaille},
  {Robitaille}, {Rod{\'o}n}, {Roy}, {Royer}, {Russeil}, {Saraceno}, {Sauvage},
  {Schilke}, {Schisano}, {Schneider}, {Schuller}, {Schulz}, {Sibthorpe},
  {Smith}, {Smith}, {Spinoglio}, {Stamatellos}, {Strafella}, {Stringfellow},
  {Sturm}, {Taylor}, {Thompson}, {Traficante}, {Tuffs}, {Umana}, {Valenziano},
  {Vavrek}, {Veneziani}, {Viti}, {Waelkens}, {Ward-Thompson}, {White},
  {Wilcock}, {Wyrowski}, {Yorke}, \& {Zhang}}]{Molinari2010}
{Molinari}, S., {Swinyard}, B., {Bally}, J., {et~al.} 2010, \aap, 518, L100

\bibitem[{{M{\"u}ller} {et~al.}(2005){M{\"u}ller}, {Schl{\"o}der}, {Stutzki},
  \& {Winnewisser}}]{Mueller2005}
{M{\"u}ller}, H. S.~P., {Schl{\"o}der}, F., {Stutzki}, J., \& {Winnewisser}, G.
  2005, Journal of Molecular Structure, 742, 215

\bibitem[{{Nagai} {et~al.}(1998){Nagai}, {Inutsuka}, \& {Miyama}}]{Nagai1998}
{Nagai}, T., {Inutsuka}, S.-i., \& {Miyama}, S.~M. 1998, \apj, 506, 306

\bibitem[{{Nguyen-Lu'o'ng} {et~al.}(2013){Nguyen-Lu'o'ng}, {Motte}, {Carlhoff},
  {Louvet}, {Lesaffre}, {Schilke}, {Hill}, {Hennemann}, {Gusdorf}, {Didelon},
  {Schneider}, {Bontemps}, {Duarte-Cabral}, {Menten}, {Martin}, {Wyrowski},
  {Bendo}, {Roussel}, {Bernard}, {Bronfman}, {Henning}, {Kramer}, \&
  {Heitsch}}]{NguyenLuong2013}
{Nguyen-Lu'o'ng}, Q., {Motte}, F., {Carlhoff}, P., {et~al.} 2013, \apj, 775, 88

\bibitem[{{Padoan} {et~al.}(2001){Padoan}, {Juvela}, {Goodman}, \&
  {Nordlund}}]{Padoan2001}
{Padoan}, P., {Juvela}, M., {Goodman}, A.~A., \& {Nordlund}, {\r{A}}. 2001,
  \apj, 553, 227

\bibitem[{{Palmeirim} {et~al.}(2013){Palmeirim}, {Andr{\'e}}, {Kirk},
  {Ward-Thompson}, {Arzoumanian}, {K{\"o}nyves}, {Didelon}, {Schneider},
  {Benedettini}, {Bontemps}, {Di Francesco}, {Elia}, {Griffin}, {Hennemann},
  {Hill}, {Martin}, {Men'shchikov}, {Molinari}, {Motte}, {Nguyen Luong},
  {Nutter}, {Peretto}, {Pezzuto}, {Roy}, {Rygl}, {Spinoglio}, \&
  {White}}]{Palmeirim2013}
{Palmeirim}, P., {Andr{\'e}}, P., {Kirk}, J., {et~al.} 2013, \aap, 550, A38

\bibitem[{{Park} \& {Ryu}(2019)}]{Park2019}
{Park}, J. \& {Ryu}, D. 2019, \apj, 875, 2

\bibitem[{{Pecaut} \& {Mamajek}(2013)}]{Pecaut2013}
{Pecaut}, M.~J. \& {Mamajek}, E.~E. 2013, \apjs, 208, 9

\bibitem[{{Planck Collaboration} {et~al.}(2016){Planck Collaboration}, {Ade},
  {Aghanim}, {Alves}, {Arnaud}, {Arzoumanian}, {Aumont}, {Baccigalupi},
  {Banday}, {Barreiro}, {Bartolo}, {Battaner}, {Benabed}, {Benoit-L{\'e}vy},
  {Bernard}, {Bern{\'e}}, {Bersanelli}, {Bielewicz}, {Bonaldi}, {Bonavera},
  {Bond}, {Borrill}, {Bouchet}, {Boulanger}, {Bracco}, {Burigana}, {Calabrese},
  {Cardoso}, {Catalano}, {Chamballu}, {Chiang}, {Christensen}, {Clements},
  {Colombi}, {Colombo}, {Combet}, {Couchot}, {Crill}, {Curto}, {Cuttaia},
  {Danese}, {Davies}, {Davis}, {de Bernardis}, {de Rosa}, {de Zotti},
  {Delabrouille}, {Dickinson}, {Diego}, {Donzelli}, {Dor{\'e}}, {Douspis},
  {Ducout}, {Dupac}, {Elsner}, {En{\ss}lin}, {Eriksen}, {Falgarone},
  {Ferri{\`e}re}, {Finelli}, {Forni}, {Frailis}, {Fraisse}, {Franceschi},
  {Frejsel}, {Galeotta}, {Galli}, {Ganga}, {Ghosh}, {Giard},
  {Giraud-H{\'e}raud}, {Gjerl{\o}w}, {Gonz{\'a}lez-Nuevo}, {G{\'o}rski},
  {Gregorio}, {Gruppuso}, {Guillet}, {Hansen}, {Hanson}, {Harrison},
  {Hern{\'a}ndez-Monteagudo}, {Herranz}, {Hildebrandt}, {Hivon}, {Hobson},
  {Holmes}, {Huffenberger}, {Hurier}, {Jaffe}, {Jaffe}, {Jones}, {Juvela},
  {Keskitalo}, {Kisner}, {Knoche}, {Kunz}, {Kurki-Suonio}, {Lagache},
  {Lamarre}, {Lasenby}, {Lawrence}, {Leonardi}, {Levrier}, {Liguori}, {Lilje},
  {Linden-V{\o}rnle}, {L{\'o}pez-Caniego}, {Lubin}, {Mac{\'{\i}}as-P{\'e}rez},
  {Maffei}, {Mandolesi}, {Mangilli}, {Maris}, {Martin},
  {Mart{\'{\i}}nez-Gonz{\'a}lez}, {Masi}, {Matarrese}, {Mazzotta},
  {Melchiorri}, {Mendes}, {Mennella}, {Migliaccio}, {Mitra},
  {Miville-Desch{\^e}nes}, {Moneti}, {Montier}, {Morgante}, {Mortlock},
  {Munshi}, {Murphy}, {Naselsky}, {Nati}, {Natoli}, {N{\o}rgaard-Nielsen},
  {Noviello}, {Novikov}, {Novikov}, {Oppermann}, {Pagano}, {Pajot}, {Paladini},
  {Paoletti}, {Pasian}, {Perrotta}, {Pettorino}, {Piacentini}, {Piat},
  {Pierpaoli}, {Pietrobon}, {Plaszczynski}, {Pointecouteau}, {Polenta},
  {Pratt}, {Puget}, {Rachen}, {Rebolo}, {Reinecke}, {Remazeilles}, {Renault},
  {Renzi}, {Ricciardi}, {Ristorcelli}, {Rocha}, {Rosset}, {Rossetti},
  {Roudier}, {Rubi{\~n}o-Mart{\'{\i}}n}, {Rusholme}, {Sandri}, {Savelainen},
  {Savini}, {Scott}, {Soler}, {Stolyarov}, {Sutton}, {Suur-Uski}, {Sygnet},
  {Tauber}, {Terenzi}, {Toffolatti}, {Tomasi}, {Tristram}, {Tucci},
  {Valenziano}, {Valiviita}, {Van Tent}, {Vielva}, {Villa}, {Wade}, {Wandelt},
  {Yvon}, {Zacchei}, \& {Zonca}}]{PlanckArzoumanian2016}
{Planck Collaboration}, {Ade}, P.~A.~R., {Aghanim}, N., {et~al.} 2016, \aap,
  586, A136

\bibitem[{{Polychroni} {et~al.}(2013){Polychroni}, {Schisano}, {Elia}, {Roy},
  {Molinari}, {Martin}, {Andr{\'e}}, {Turrini}, {Rygl}, {Di Francesco},
  {Benedettini}, {Busquet}, {di Giorgio}, {Pestalozzi}, {Pezzuto},
  {Arzoumanian}, {Bontemps}, {Hennemann}, {Hill}, {K{\"o}nyves},
  {Men'shchikov}, {Motte}, {Nguyen-Luong}, {Peretto}, {Schneider}, \&
  {White}}]{Polychroni2013}
{Polychroni}, D., {Schisano}, E., {Elia}, D., {et~al.} 2013, \apjl, 777, L33

\bibitem[{{Pon} {et~al.}(2012){Pon}, {Johnstone}, \& {Kaufman}}]{Pon2012}
{Pon}, A., {Johnstone}, D., \& {Kaufman}, M.~J. 2012, \apj, 748, 25

\bibitem[{{Pon} {et~al.}(2014){Pon}, {Johnstone}, {Kaufman}, {Caselli}, \&
  {Plume}}]{Pon2014}
{Pon}, A., {Johnstone}, D., {Kaufman}, M.~J., {Caselli}, P., \& {Plume}, R.
  2014, \mnras, 445, 1508

\bibitem[{{Pound} \& {Wolfire}(2008)}]{Pound2008}
{Pound}, M.~W. \& {Wolfire}, M.~G. 2008, in Astronomical Society of the Pacific
  Conference Series, Vol. 394, Astronomical Data Analysis Software and Systems
  XVII, ed. R.~W. {Argyle}, P.~S. {Bunclark}, \& J.~R. {Lewis}, 654

\bibitem[{{Rayner} {et~al.}(2017){Rayner}, {Griffin}, {Schneider}, {Motte},
  {K{\"o}nyves}, {Andr{\'e}}, {Di Francesco}, {Didelon}, {Pattle},
  {Ward-Thompson}, {Anderson}, {Benedettini}, {Bernard}, {Bontemps}, {Elia},
  {Fuente}, {Hennemann}, {Hill}, {Kirk}, {Marsh}, {Men'shchikov}, {Nguyen
  Luong}, {Peretto}, {Pezzuto}, {Rivera-Ingraham}, {Roy}, {Rygl},
  {S{\'a}nchez-Monge}, {Spinoglio}, {Tig{\'e}}, {Trevi{\~n}o-Morales}, \&
  {White}}]{Rayner2017}
{Rayner}, T.~S.~M., {Griffin}, M.~J., {Schneider}, N., {et~al.} 2017, \aap,
  607, A22

\bibitem[{{Risacher} {et~al.}(2018){Risacher}, {G{\"u}sten}, {Stutzki},
  {H{\"u}bers}, {Aladro}, {Bell}, {Buchbender}, {B{\"u}chel}, {Csengeri},
  {Duran}, {Graf}, {Higgins}, {Honingh}, {Jacobs}, {Justen}, {Klein},
  {Mertens}, {Okada}, {Parikka}, {P{\"u}tz}, {Reyes}, {Richter}, {Ricken},
  {Riquelme}, {Rothbart}, {Schneider}, {Simon}, {Wienold}, {Wiesemeyer},
  {Ziebart}, {Fusco}, {Rosner}, \& {Wohler}}]{Risacher2018}
{Risacher}, C., {G{\"u}sten}, R., {Stutzki}, J., {et~al.} 2018, Journal of
  Astronomical Instrumentation, 7, 1840014

\bibitem[{{R{\"o}llig} {et~al.}(2007){R{\"o}llig}, {Abel}, {Bell}, {Bensch},
  {Black}, {Ferland}, {Jonkheid}, {Kamp}, {Kaufman}, {Le Bourlot}, {Le Petit},
  {Meijerink}, {Morata}, {Ossenkopf}, {Roueff}, {Shaw}, {Spaans}, {Sternberg},
  {Stutzki}, {Thi}, {van Dishoeck}, {van Hoof}, {Viti}, \&
  {Wolfire}}]{Roellig2007}
{R{\"o}llig}, M., {Abel}, N.~P., {Bell}, T., {et~al.} 2007, \aap, 467, 187

\bibitem[{{R{\"o}llig} \& {Ossenkopf}(2013)}]{Roellig2013}
{R{\"o}llig}, M. \& {Ossenkopf}, V. 2013, \aap, 550, A56

\bibitem[{{Schilke} {et~al.}(1997){Schilke}, {Walmsley}, {Pineau des Forets},
  \& {Flower}}]{Schilke1997}
{Schilke}, P., {Walmsley}, C.~M., {Pineau des Forets}, G., \& {Flower}, D.~R.
  1997, \aap, 321, 293

\bibitem[{{Schisano} {et~al.}(2014){Schisano}, {Rygl}, {Molinari}, {Busquet},
  {Elia}, {Pestalozzi}, {Polychroni}, {Billot}, {Carey}, {Paladini},
  {Noriega-Crespo}, {Moore}, {Plume}, {Glover}, \&
  {V{\'a}zquez-Semadeni}}]{Schisano2014}
{Schisano}, E., {Rygl}, K.~L.~J., {Molinari}, S., {et~al.} 2014, \apj, 791, 27

\bibitem[{{Schneider} {et~al.}(2016){Schneider}, {Bontemps}, {Motte},
  {Blazere}, {Andr{\'e}}, {Anderson}, {Arzoumanian}, {Comer{\'o}n}, {Didelon},
  {Di Francesco}, {Duarte-Cabral}, {Guarcello}, {Hennemann}, {Hill},
  {K{\"o}nyves}, {Marston}, {Minier}, {Rygl}, {R{\"o}llig}, {Roy}, {Spinoglio},
  {Tremblin}, {White}, \& {Wright}}]{Schneider2016}
{Schneider}, N., {Bontemps}, S., {Motte}, F., {et~al.} 2016, \aap, 591, A40

\bibitem[{{Schneider} {et~al.}(2011){Schneider}, {Bontemps}, {Simon},
  {Ossenkopf}, {Federrath}, {Klessen}, {Motte}, {Andr{\'e}}, {Stutzki}, \&
  {Brunt}}]{Schneider2011}
{Schneider}, N., {Bontemps}, S., {Simon}, R., {et~al.} 2011, \aap, 529, A1

\bibitem[{{Schneider} {et~al.}(2012){Schneider}, {Csengeri}, {Hennemann},
  {Motte}, {Didelon}, {Federrath}, {Bontemps}, {Di Francesco}, {Arzoumanian},
  {Minier}, {Andr{\'e}}, {Hill}, {Zavagno}, {Nguyen-Luong}, {Attard},
  {Bernard}, {Elia}, {Fallscheer}, {Griffin}, {Kirk}, {Klessen}, {K{\"o}nyves},
  {Martin}, {Men'shchikov}, {Palmeirim}, {Peretto}, {Pestalozzi}, {Russeil},
  {Sadavoy}, {Sousbie}, {Testi}, {Tremblin}, {Ward-Thompson}, \&
  {White}}]{Schneider2012}
{Schneider}, N., {Csengeri}, T., {Hennemann}, M., {et~al.} 2012, \aap, 540, L11

\bibitem[{{Smith} {et~al.}(2016){Smith}, {Glover}, {Klessen}, \&
  {Fuller}}]{Smith2016}
{Smith}, R.~J., {Glover}, S.~C.~O., {Klessen}, R.~S., \& {Fuller}, G.~A. 2016,
  \mnras, 455, 3640

\bibitem[{{Sz{\H{u}}cs} {et~al.}(2014){Sz{\H{u}}cs}, {Glover}, \&
  {Klessen}}]{Szucs2014}
{Sz{\H{u}}cs}, L., {Glover}, S. C.~O., \& {Klessen}, R.~S. 2014, \mnras, 445,
  4055

\bibitem[{{Tielens} \& {Hollenbach}(1985)}]{Tielens1985}
{Tielens}, A.~G.~G.~M. \& {Hollenbach}, D. 1985, \apj, 291, 722

\bibitem[{{Tritsis} \& {Tassis}(2018)}]{Tritsis2018}
{Tritsis}, A. \& {Tassis}, K. 2018, Science, 360, 635

\bibitem[{{van der Tak} {et~al.}(2007){van der Tak}, {Black}, {Sch{\"o}ier},
  {Jansen}, \& {van Dishoeck}}]{vanderTak2007}
{van der Tak}, F.~F.~S., {Black}, J.~H., {Sch{\"o}ier}, F.~L., {Jansen}, D.~J.,
  \& {van Dishoeck}, E.~F. 2007, \aap, 468, 627

\bibitem[{{V{\'a}zquez-Semadeni} {et~al.}(2019){V{\'a}zquez-Semadeni}, {Palau},
  {Ballesteros-Paredes}, {G{\'o}mez}, \&
  {Zamora-Avil{\'e}s}}]{VazquezSemadeni2019}
{V{\'a}zquez-Semadeni}, E., {Palau}, A., {Ballesteros-Paredes}, J.,
  {G{\'o}mez}, G.~C., \& {Zamora-Avil{\'e}s}, M. 2019, \mnras, 490, 3061

\bibitem[{{Vilas-Boas} {et~al.}(1994){Vilas-Boas}, {Myers}, \&
  {Fuller}}]{VilasBoas1994}
{Vilas-Boas}, J.~W.~S., {Myers}, P.~C., \& {Fuller}, G.~A. 1994, \apj, 433, 96

\bibitem[{{Visser} {et~al.}(2009){Visser}, {van Dishoeck}, \&
  {Black}}]{Visser2009}
{Visser}, R., {van Dishoeck}, E.~F., \& {Black}, J.~H. 2009, \aap, 503, 323

\bibitem[{{Wenger} {et~al.}(2000){Wenger}, {Ochsenbein}, {Egret}, {Dubois},
  {Bonnarel}, {Borde}, {Genova}, {Jasniewicz}, {Lalo{\"e}}, {Lesteven}, \&
  {Monier}}]{Wenger2000}
{Wenger}, M., {Ochsenbein}, F., {Egret}, D., {et~al.} 2000, \aaps, 143, 9

\bibitem[{{Whitworth} \& {Jaffa}(2018)}]{Whitworth2018}
{Whitworth}, A.~P. \& {Jaffa}, S.~E. 2018, \aap, 611, A20

\end{thebibliography}
%

\begin{appendix}

\section{SOFIA \OI\ and \CII\ observations} \label{SOFIA} 

\begin{figure*}
\begin{center}
\includegraphics[width=0.45\hsize]{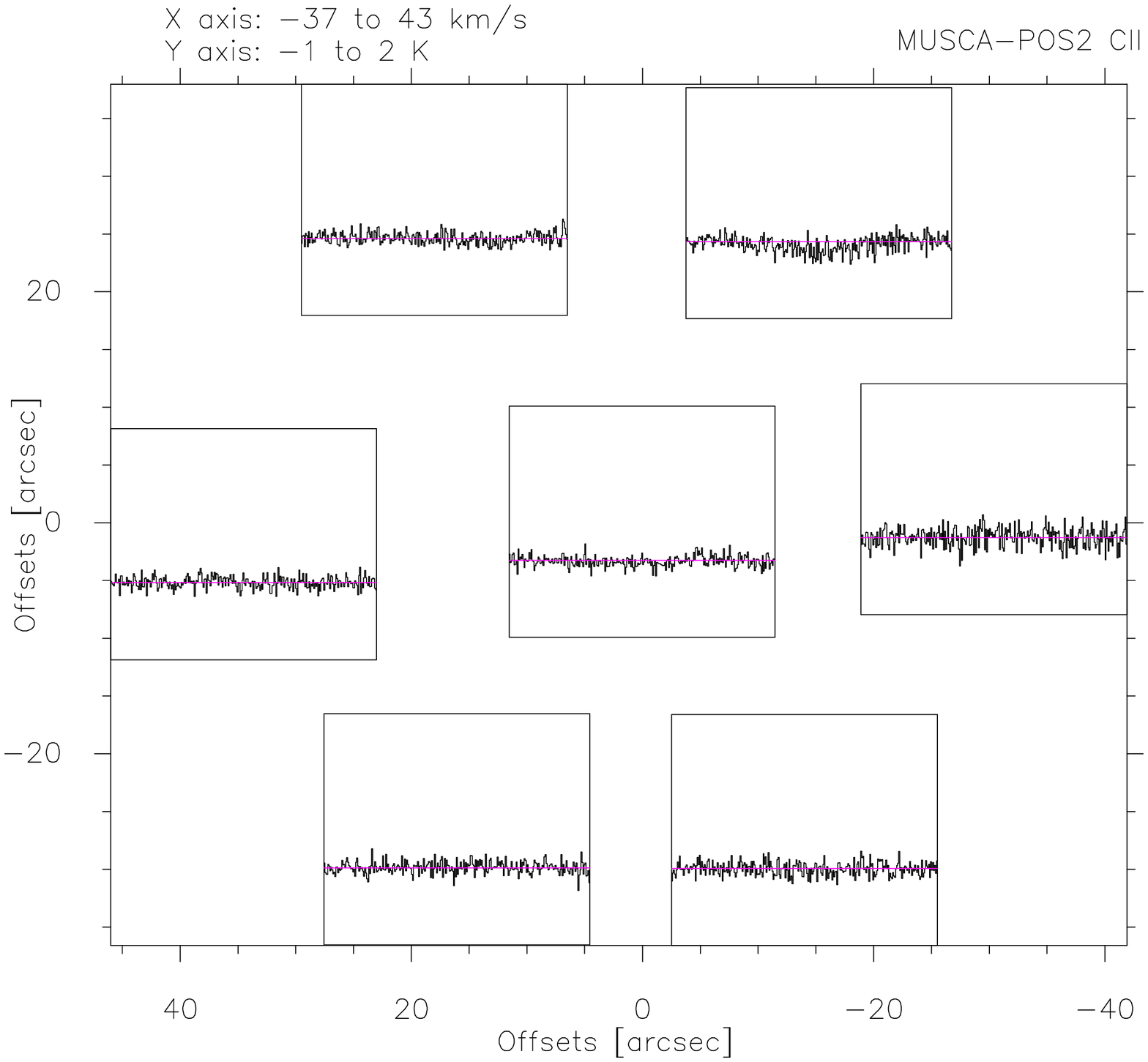}
\includegraphics[width=0.45\hsize]{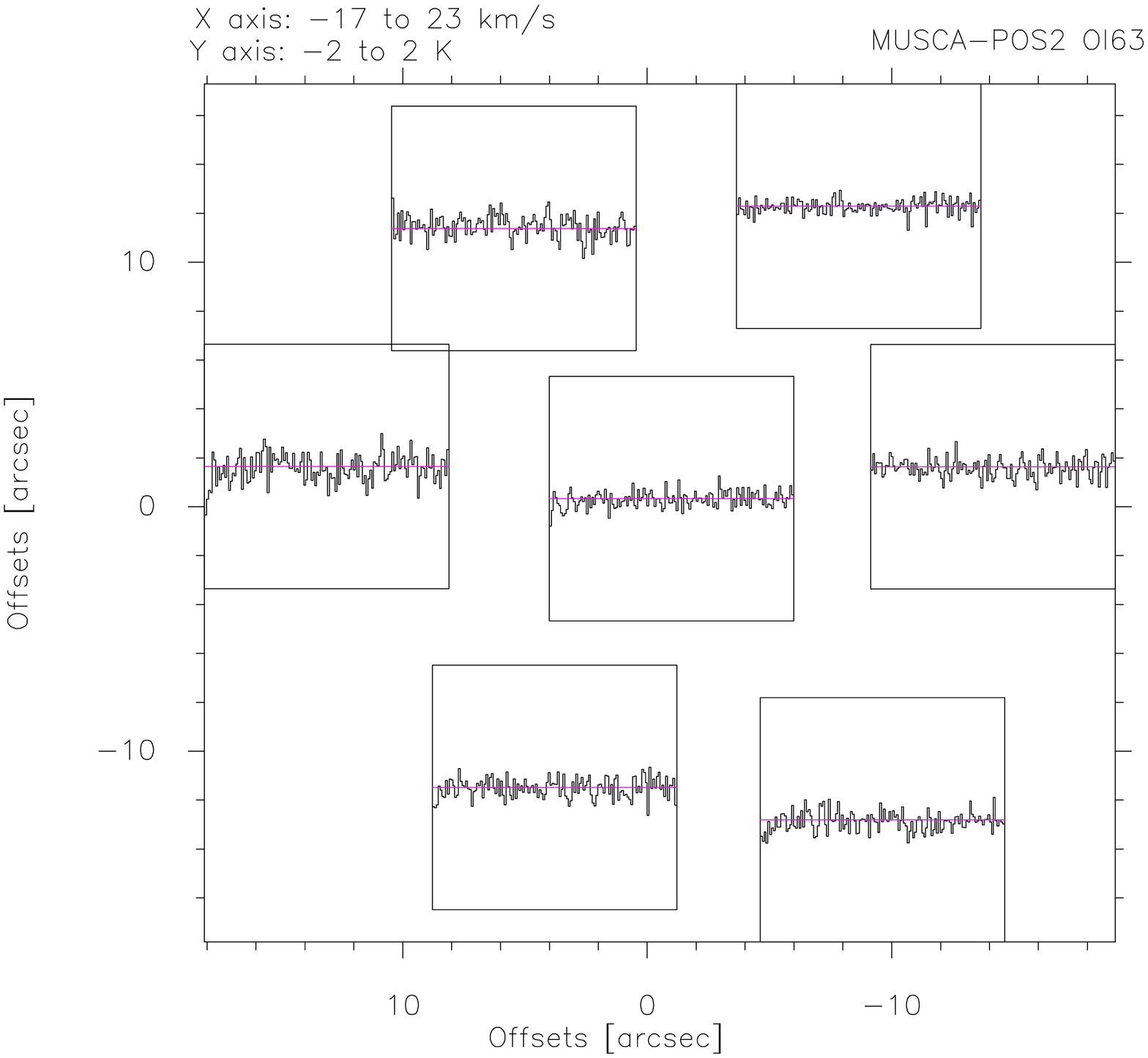}
\end{center}
\caption{\textbf{Left}: Observed pixels for the \CII\ 158 $\mu$m line with the SOFIA telescope with a spectral resolution of 0.3 km s$^{-1}$. \textbf{Right}: Observed pixels for the
  \OI\ 63 $\mu$m line.}
\label{pixelsCII-OI}
\end{figure*}

\begin{figure}
\includegraphics[width=\hsize]{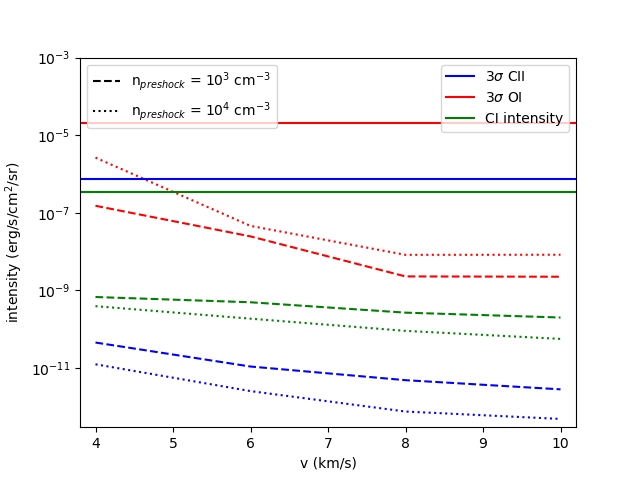}
\caption{Intensity of \CII\ (blue), \OI\ (red), and \CI\ (green) predicted for J-type shock models with a preshock density of n = 10$^{3}$ cm$^{-3}$ and n = 10$^{4}$ cm$^{-3}$ by the Paris-Durham code. The horizontal lines indicate the observed \CI\ intensity and 3 $\sigma$ upper limits for \CII\ and \OI. This shows that the \CII\ and \OI\ non-detections in Musca are consistent with shocks and that the \CI\ brightness from the shock is not brighter than the total observed \CI\ brightness in Musca.}
\label{OICIICIparisdurham}
\end{figure}

The individual pixels of the upGREAT LFA and HFA, covering the
\CII\ 158 $\mu$m line and the \OI\ 63 $\mu$m line, respectively, are
displayed in Fig. \ref{pixelsCII-OI}. Both lines, \CII\ and \OI\ are
not detected above an rms (depending on the pixel) of 0.06 and 0.13 K
for the LFA, and 0.11 and 0.21 K for the HFA.\\\\
\OI\ is a possible shock tracer, but its non detection does not argue against a possible shock excitation, see Fig. \ref{OICIICIparisdurham}, because of the difficulty to excite \OI\ emission. It either needs high-velocity shocks \citep{Hollenbach1989}, an irradiated lower velocity shock \citep{Lesaffre2013,Godard2019} or dense J/CJ-type shocks \citep{Flower2010,Gusdorf2017}. Simulations of massive molecular cloud formation through colliding
flows predict an \OI\ brightness around 0.01 - 0.05 K km
s$^{-1}$ \citep{Bisbas2017}. Yet, such predicted brightnesses are highly uncertain because of the  unconstrained atomic oxygen abundance in the ISM \citep[e.g.][]{Meyer1998,Cartledge2004,Jenkins2009,Hincelin2011}.

\section{A two layer multi-component model } \label{fittingAppendix}
\begin{figure*}
\begin{center}
\includegraphics[width=0.45\hsize]{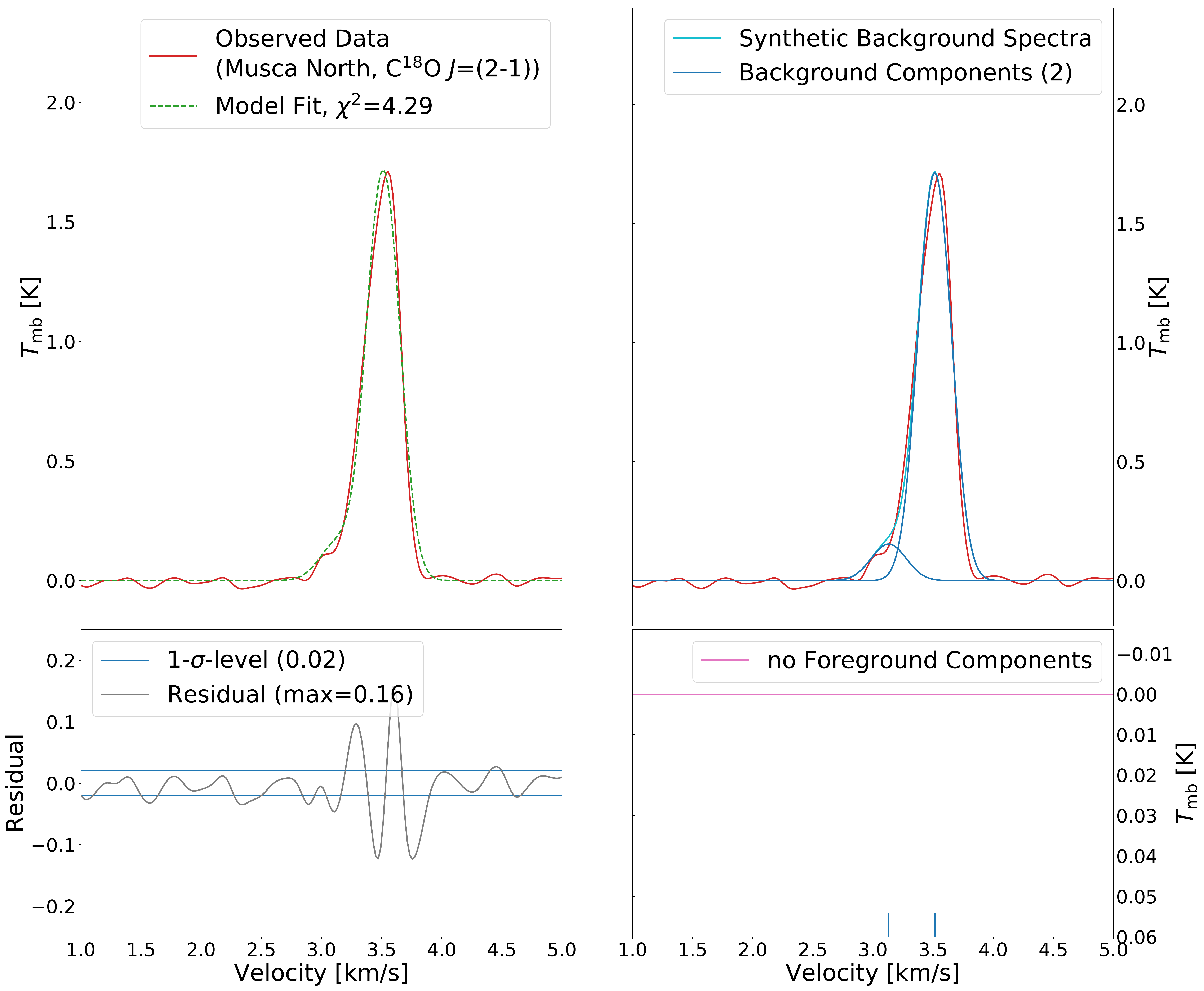}
\includegraphics[width=0.45\hsize]{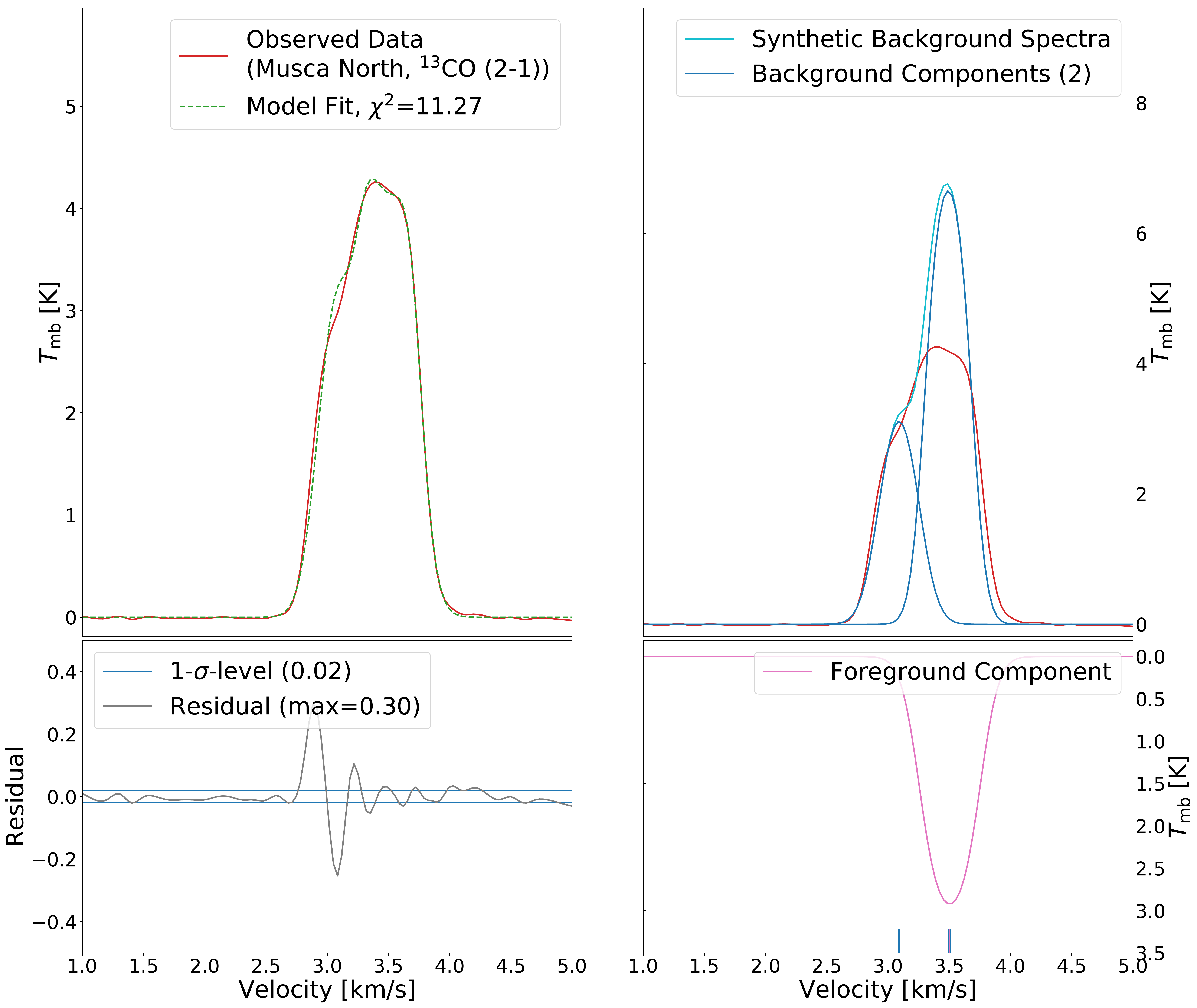}
\includegraphics[width=0.45\hsize]{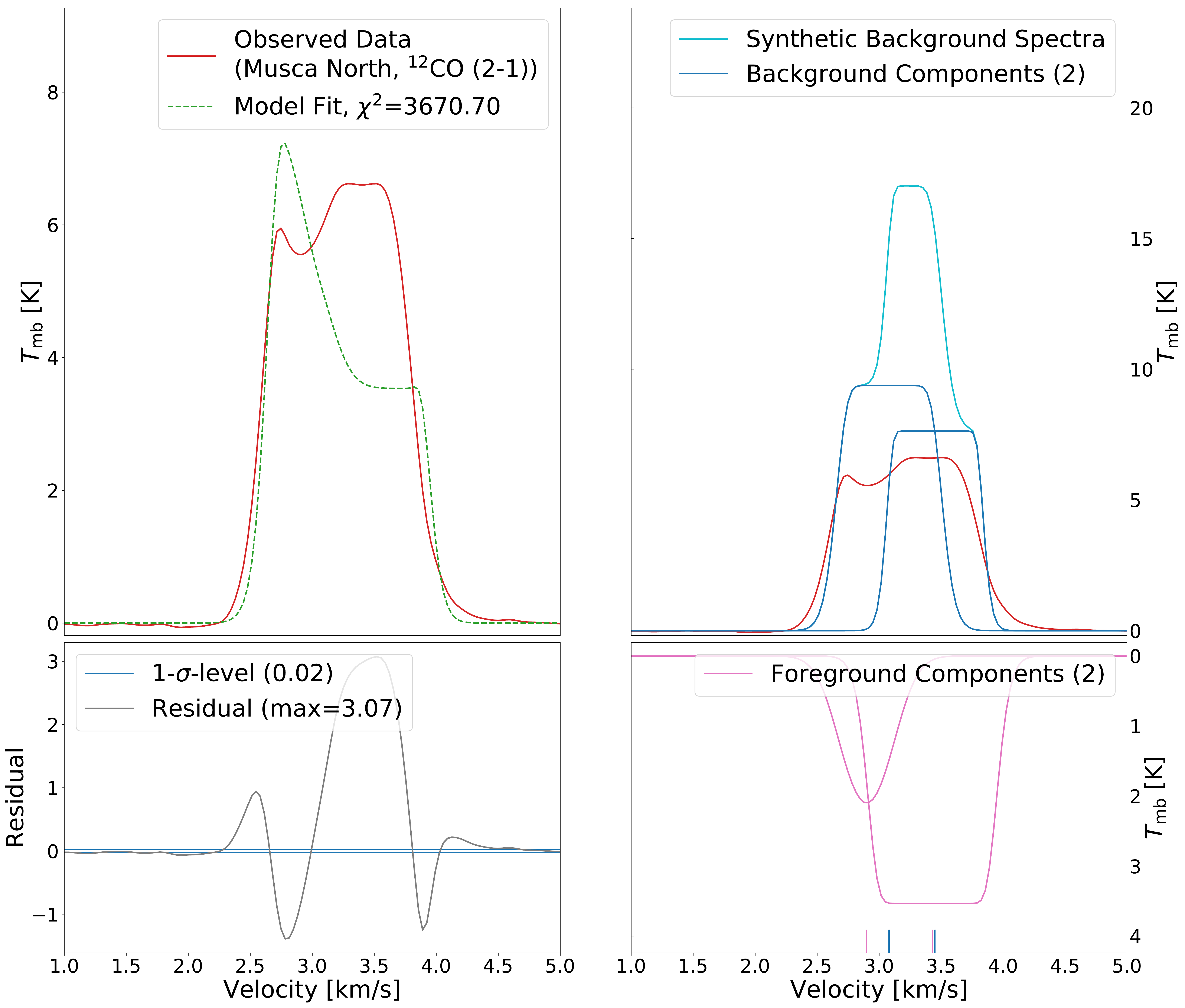}
\includegraphics[width=0.45\hsize]{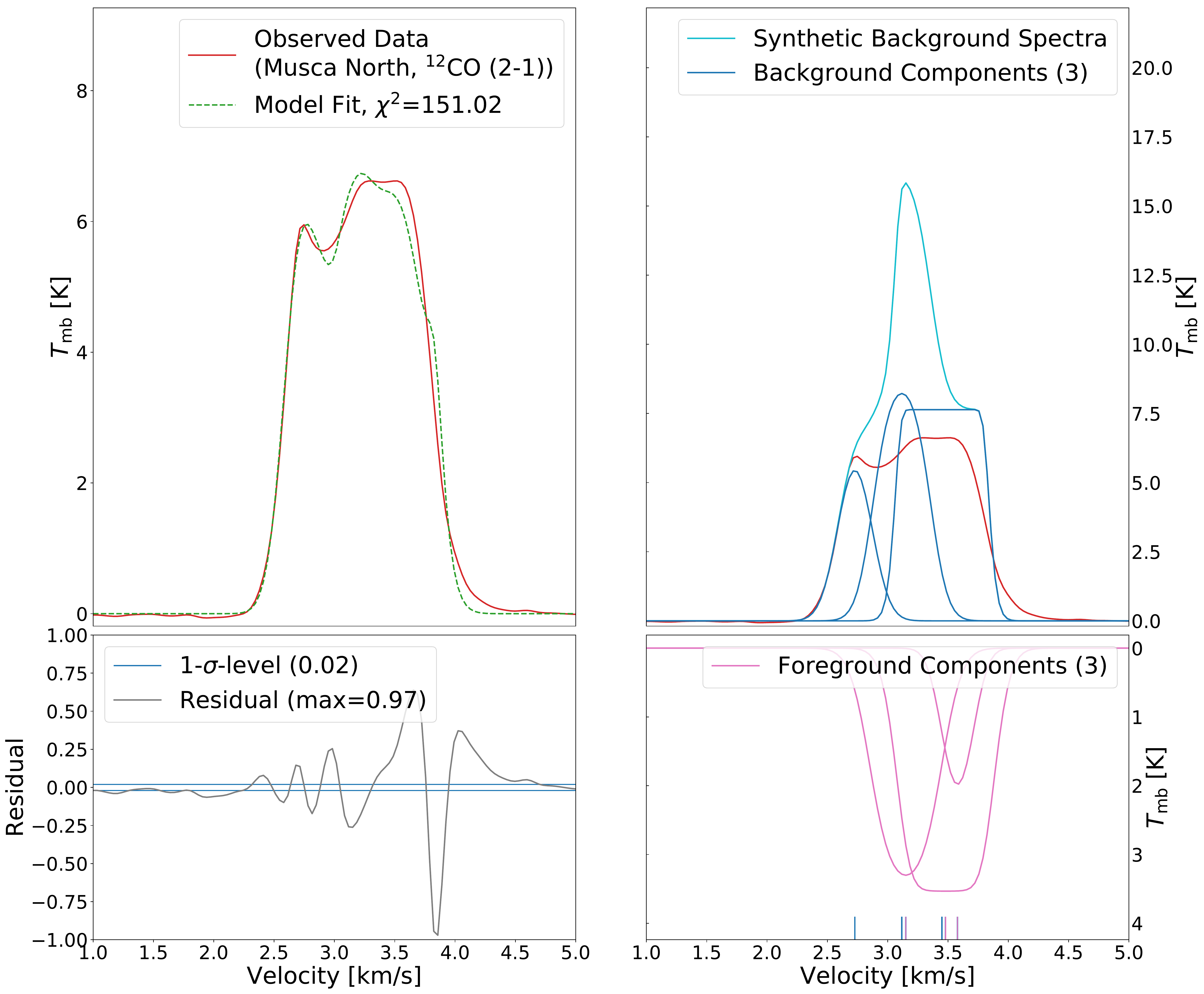}
\end{center}
\caption{Results from a two-layer, multi-component fit to the observed CO(2-1) isotopes at the northern Musca position. The upper two panels and the lower left panel show a fit with two velocity components derived from the optically thin C$^{18}$O(2-1) emission line, the lower right panel shows the best model fit for the observed $^{12}$CO(2-1) line.  In each panel, the upper left sub-panel shows the observed data as a red line and the model fit as a green dashed line. The upper right sub-panel displays the individual components, the lower left sub-panel shows the residuals and the lower right sub-panel gives information on the number and contribution of the foreground. It becomes obvious that two components can be fitted for C$^{18}$O(2-1) and $^{13}$CO(2-1), but are insufficient for the $^{12}$CO(2-1) line.}
\label{cofittingnorth}
\end{figure*}

To investigate possible effects of self-absorption, we apply a multi-component dual layer
model \citep{Guevara2020} to the observations of the CO(2-1) isotopologues. 
This LTE model fits the radiative transfer equations for multiple components distributed in two layers. 
The assumption of LTE for low-J CO lines is a justified approximation, as discussed in for example \citet{Goldsmith1999,Liszt2006}. 
\citet{vanderTak2007} also show that the LTE model is mostly valid for the ground state CO(1-0) transition and deviations
from this approximation increase with the quantum number of the upper level. Here, we only use the CO(2-1) isotologue transitions.  The fitted radiative transfer equation is given by

\begin{equation}  \label{eq:tmb} 
\begin{split}
T_{\mathrm{mb}}(\rm {v}) = & \left[ \sum_{i_\mathrm{bg}} \mathcal{J}_{\nu}(T_{\mathrm{ex},i_\mathrm{bg}}) \, \left( 1-e^{-\tau_{i_\mathrm{bg}}(\rm {v})}\right) \right] e^{-\sum_{i_\mathrm{fg}} \tau_{i_\mathrm{fg}}(\rm {v})} + \\
& \sum_{i_\mathrm{fg}} \mathcal{J}_{\nu}(T_{\mathrm{ex},i_\mathrm{fg}}) \, \left( 1-e^{-\tau_{i_\mathrm{fg}}(\rm {v})}\right),  \\
\end{split}
\end{equation} 

\noindent
with $\mathcal{J}_{\nu}(T_{\mathrm{ex}}$) the equivalent brightness temperature of blackbody emission at a temperature $T_{\mathrm{ex}}$:

\begin{equation}
	 \mathcal{J}_{\nu}(T_{\mathrm{ex}}) = \frac{h\nu}{k_B}\frac{1}{e^{T_0/T_{\mathrm{ex}}}-1},
\end{equation}
 
 \noindent
with the equivalent temperature of the excited level $T_0=h\nu/k_B $  and $\nu$ the transition frequency.  
The single components of the model are represented by Gaussian profiles, thus the velocity dependent optical depth of every component is given by:

\begin{equation}
	\tau(v)=\tau_0e^{-4\ln 2~\left(\frac{\rm {v}-{\rm v}_0}{w}\right)^2},
\end{equation}

with ${\rm v}_0$ the central (LSR) velocity of the component and $w$ 
the line width (FWHM). 
 
Each of our components (background and foreground alike) are characterized by four
quantities: excitation temperature, optical depth, position (LSR
velocity), and width (FWHM). For a given number of components the 
model fit converges towards the best values for each
parameter in a given range. The line position and its width are confined by the 
observed line profile, thus the excitation temperature and optical depth remain free parameter.  
The combination of $T_{\mathrm{ex}}$ and $\tau$ is degenerate, that is, 
higher excitation temperatures go together with lower optical depths and
vice versa (and then give a similar fit). However, we can to first order 
constrain the gas temperature using the {\sl Herschel} dust temperature $T_d$. 

The Musca filament (in particular in the crest),
is most likely heated by cosmic rays because neither thermal (low densities) nor radiative heating
(low FUV field) dominates. The main cooling of the gas is happening
via the low-J CO lines. Because of the low densities, freeze-out of
CO is probably not so significant as it is for high densities.
Standard dust models \citep{Goldsmith2001} predict a difference
between gas and dust temperature (with T$_g>$T$_d$),
while recent studies \citep{Ivlev2019}, considering also the grain
size distribution, anticipate a less strong difference between gas
and dust temperatures.  We thus use as a starting value the dust
temperatures derived from {\sl Herschel} shown in
Fig. \ref{singleSpectraCO43} and apply a gas excitation
temperature in the range of 10 to 13 K for the filament crest
component and of T $>$ 13 K for the shoulder+blueshifted component. 
Therefore, only the optical depth remains as a free parameter. 

The two layer multi-component model is applied to the northern position 
(the results for the southern position are similar) iteratively for the three transition lines:
$\mathrm{C^{18}O}$(2-1), $\mathrm{^{13}CO}$(2-1) and  $\mathrm{^{12}CO}$(2-1), as shown in Fig.  \ref{cofittingnorth}.

We start with the C$^{18}$O(2-1) line where we expect no absorption due to the foreground. 
This way we determine the background components, detected with this line, which are simply scaled up by the natural abundance ratio
for the two remaining molecular carbon lines. The physical parameter of each component obtained from the model fit are summarized in Table \ref{tabfittedSpectra} and plotted in the upper left panel of Fig. \ref{cofittingnorth}. This fit constrains the position and optical depth of the  background component. 

We then continue to model the $^{13}$CO(2-1) line profile. Because the background components are here constrained by the C$^{18}$O(2-1) fit, we scale up the background components by the $\mathrm{^{13}CO/C^{18}O}\sim 7$ ratio. We do not strictly fix the parameter obtained from the C$^{18}$O(2-1) model fit, but allow the model to find the best fit in a small range around the given parameter. 

The scaled up background emission overshoots the observed line at v$\sim$ 3.5 km s$^{-1}$. To model the observed line intensities, we add a cold foreground layer with T$<$10. The foreground component absorbs the overshooting emission
, see the upper right panel of Fig. \ref{cofittingnorth}. Note, however, that it is unlikely that we have a cooler layer that $'$wraps around$'$ the filament (which we consider a cylinder) because  the {\sl Herschel} temperature map clearly shows an increasing temperature when going outwards from the crest, see Fig. \ref{singleSpectraCO43}. The only possible way that one could have cool foreground emission that can absorb emission, is by considering a clumpy medium that is unresolved by the \textit{Herschel} beam. In this unresolved clumpy medium, located in the foreground of the line of sight, one then would need mixed relatively cold ($<$ 10 K) and warm gas ($>$ 15 K) which gives rise to a steadily increasing average temperature in the \textit{Herschel} beam when moving outwards from the filament crest.

The second blueshifted component only matches the observed emission with an increased $\mathrm{^{13}CO/C^{18}O}$ abundance ratio, see Table \ref{tabfittedSpectra}. The abundance ratio between the molecules is not a very well defined value. 
The processes of selective photodissociation and carbon isotopic fractionation can modify the relative abundances
of CO isotopologues \citep[e.g.][]{Visser2009,Liszt2017}. In particular in more diffuse regions, the fractionation via the exchange reaction between $^{13}$C$^+$ and $^{12}$CO leads to an enhancement of the $^{13}$CO abundance \citep[e.g.][]{Liszt2012,Roellig2013,Szucs2014}. 
In addition - though probably less important for Musca because of the low FUV field - the more abundant $^{12}$CO and $^{13}$CO isotopologues shield themselves from the destructive effect of FUV photons more efficiently than the less
abundant C$^{18}$O isotopologue because the photodissociation of CO is governed by line absorption. In this context we note that \citet{Hacar2016} found indications of strong fractionation leading to a significant increase of the $^{13}$CO/C$^{18}$O abundance ratio above the standard value of 7 at A$_{\rm V} <$ 3. Further analysis of this abundance ratio in our companion paper on the Musca filament confirms this result presented in \citet{Hacar2016}.



In the previous two model fits, we determined the background and foreground components which we now scale up by the abundance ratio $\mathrm{^{12}CO/^{13}CO}\sim 60$. Note that this value can be significantly lower because of fractionation (see above). 
The resulting fit is shown in the left lower panel of Fig. \ref{cofittingnorth}. Due to the large optical depths, we observe flat-topped spectra in the background and foreground layer. Since the observed intensities do not fit the natural carbon abundance ratio for the second component we only fix the first component and allowed the model to vary over a range of values in the second component to find the best possible solution. In addition we included a foreground component, which would only be observable in the $^{12}$CO isotope, to remove the emission excess in the blueshifted wing and possibly reproduce the bump located in the blueshifted part of the observed $^{12}$CO spectrum. The resulting spectra show a flat spectrum at velocities with bright C$^{18}$O(2-1) emission, followed by an emission peak at the blue-shifted wing. Thus, the model results do not manage to represent the line shape, the observed bump in particular, and intensity by taking self- or foreground absorption effects of the scaled up background components derived from the isotopes, see lower left panel of Fig. \ref{cofittingnorth}.    

Since two background components do not represent the observed spectrum, we include an additional component to the model, which is only visible in the $^{12}$CO spectrum but vanishes in the noise for the weaker isotopes, see lower right panel of Fig. \ref{cofittingnorth}. To meet this criteria we need to include a background component with low optical depth, which results in an increased temperature since a low optical depth requires an increase in temperature to produce the same amount of emission. This third background component is located at v = 2.8 km s$^{-1}$, see Table \ref{tabfittedSpectra}. In addition we allow cold optically thin components in order to correct for the emission excess in the central and red-shifted part of the spectrum. As in the previous model fit we do not fix the central second component by the carbon abundance ratio, but allow it to vary in a wide range to find the best model fit to the observed data.  We thus manage to represent the spectrum well, only with three background components.


\begin{table*}
\begin{center}
\begin{tabular}{c|c|c|c|c|c|c|c|c}
\hline
\multicolumn{5}{c}{\hspace{2.9cm} \bf Background} & \multicolumn{4}{|c}{\bf Foreground}   \\
\hline
  Fit parameter     &  T$_{ex}$ & $\tau$ & v & $\Delta$v &  T$_{ex}$  & $\tau$ & v  & $\Delta$v    \\
   Units &  (K)  &  & (km s$^{-1}$) & (km s$^{-1}$) &  (K)  &  & (km s$^{-1}$) & (km s$^{-1}$) \\
 \hline
 C$^{18}$O  component 1 & 12.50 & 0.24 & 3.51 & 0.31 & - & - & - &  - \\
 C$^{18}$O  component 2 & 14.00 & 0.02 & 3.13 & 0.34 & - & - & - &  - \\
  \hline
 $^{13}$CO component 1  & 12.70 & 1.70 & 3.49 & 0.31 & 7.69 & 1.70 & 3.50 & 0.39 \\
 $^{13}$CO component 2  & 13.80 & 0.41 & 3.09 & 0.35 &  -   & -   & -   & - \\
 \hline
 $^{12}$CO component 1  & 12.35 & 90.00 & 3.45 & 0.30 & 7.80  & 7.90 & 3.48 & 0.44 \\
 $^{12}$CO component 2  & 14.80 & 1.75  & 3.12 & 0.37 & 8.13  & 2.00 & 3.15 & 0.45 \\
 $^{12}$CO component 3  & 69.00 & 0.09  & 2.73 & 0.34 & 10.00 & 0.45 & 3.58 & 0.28 \\
\hline
\end{tabular}
\caption{ Model results for the two layer multi-component fit.}
\label{tabfittedSpectra}
\end{center}
\end{table*}

\begin{figure*}
\includegraphics[width=0.5\hsize]{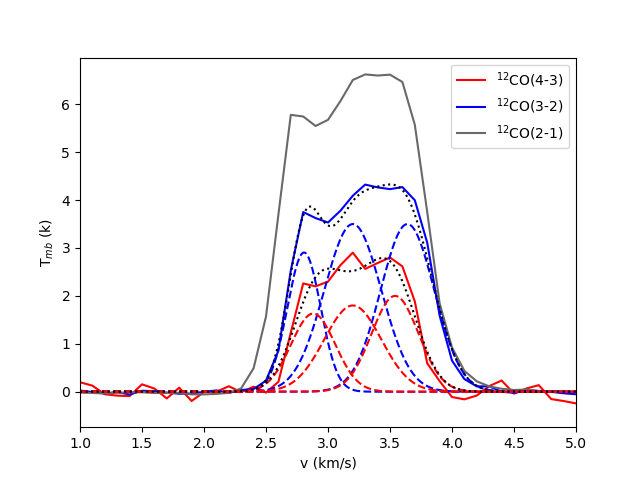}
\includegraphics[width=0.5\hsize]{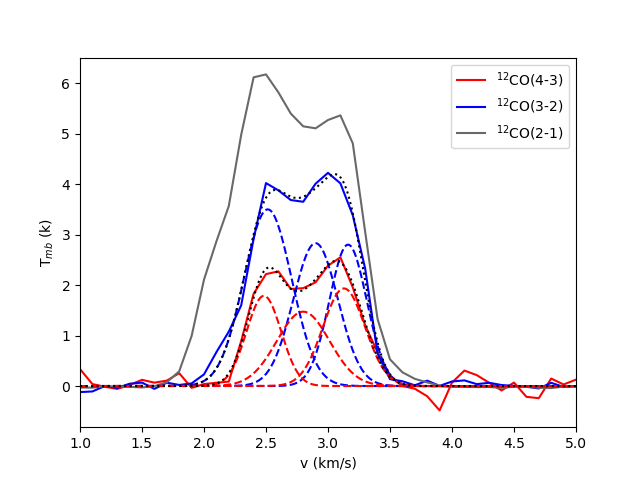}
\caption{Results of a fit of three Gaussian profiles to the $^{12}$CO(3-2) and $^{12}$CO(4-3) lines towards the crest in the northern map (left) 
and southern map (right). The solid lines are the observed data, the long-dashed lines the fit with three individual components, and the short-dashed 
lines the resulting spectrum. }
\label{fittedSpectra}
\end{figure*} 

\section{SiO non-detection}
\label{SiOapp}

The observations with the PI230 instrument on APEX towards the Musca
cloud presented in Paper I covered the SiO(5-4) line. We have presented
the detection of warm gas heated by low-velocity shocks, which makes
the observation of SiO(5-4) interesting since some molecular clouds
display extended SiO emission with relatively narrow linewidths
\citep{JimenezSerra2010,NguyenLuong2013,DuarteCabral2014}. This suggests that this is not SiO sputtered from the grains by
protostellar outflows \citep[e.g.][]{Schilke1997,Gusdorf2008a}. To explain this extended SiO emission in W43, it was considered that 1-10\% of Si was not locked in the grains in the preshock gas \citep{Louvet2016}.

%
In Musca, because of the low-velocity shocks, sputtering of SiO from the grains
seems improbable \citep{Gusdorf2008b,Louvet2016}. Upon analysing the data
after averaging over the filament crest (N $>$
3$\cdot$10$^{21}$ cm$^{-2}$), no SiO emission is detected with an rms of
0.0134 K.

Since SiO has a high critical density ($>$ 10$^{6}$ cm$^{-3}$) it
requires a non-LTE approach \citep{Csengeri2016} to study the maximal
SiO abundance in the gas phase. Therefore, we use RADEX to search for
an upper limit on the SiO column density. This is done for two
densities 10$^{3}$ and 10$^{4}$ cm$^{-3}$, assuming a minimal
temperature of 10 K and a FWHM of 0.5 km s$^{-1}$ (similar to the
C$^{18}$O linewidth). The resulting 3$\sigma$ SiO column density upper
limits are $\sim$ 10$^{14}$ and 10$^{15}$ cm$^{-2}$ for densities of 10$^{4}$
and 10$^{3}$ cm$^{-3}$ respectively.

%
%
This upper limit is observed toward the filament crest, which has a
{\sl Herschel} dust column density of $\sim$ 4$\cdot$10$^{21}$ cm$^{-2}$,
such that the upper limit obtained at a density of 10$^{4}$
cm$^{-3}$ should hold, see Paper I. Assuming all hydrogen is in
molecular form, using $\big[$Si$\big]$/$\big[$H$\big]$ =
2$\cdot$10$^{-5}$ and taking that all Si is found in the form of SiO
\citep{Louvet2016}. This poses an upper limit of 0.06 \% on the Si
abundance in the gas phase (0.6 \% when assuming n$_{H_{2}}$ =
10$^{3}$ cm$^{-3}$).\\ This indicates that in a molecular
cloud at an early stage of evolution basically all Si is locked in the grains and that one needs
 environments with a radiative or dynamic history like W43 or protostellar outflows to release SiO
in the gas phase.



\section{Mass accretion rate}
\label{appMassAccretionRate}
As we have presented the detection of a filament accretion shock, it is possible to estimate the mass accretion rate as a result of this shock using 2$\pi$R$\cdot\Sigma_{\rm shock}\cdot$t$_{\rm psc}^{-1}$. Where R is the radius that encompasses most dense gas ($\approx$ 0.2 pc), $\Sigma_{\rm shock}$ is the column density of the shock layer, and t$_{\rm psc}$ is the post-shock cooling time for which we use the expression in \citet{Whitworth2018}
\begin{equation}
{\rm t}_{\rm psc} \lesssim 1.2 {\rm Myr} \left( \frac{X_{\rm CO}}{3\cdot10^{-4}} \right)^{-1} \left( \frac{n_{H_{2},0}}{\rm cm^{-3}} \right)^{-1} \left( \frac{\rm v_{s}}{\rm km s^{-1}} \right)^{-1},
\label{tpscEq}
\end{equation}
with $n_{H_{2},0}$ the pre-shock density and v$_{\rm s}$ the shock velocity.\\
To calculate $\Sigma_{\rm shock}$ we start from the $\Sigma_{\rm ^{12}CO}$ column density of the RADEX models that manage to reach the observed ratios in Sec. \ref{sec: warmGasSection}, which is 9$\cdot$10$^{14}$ cm$^{-2}$. $\Sigma_{\rm shock}$ is then given by $\Sigma_{\rm ^{12}CO}$/X$_{\rm CO}$. This X$_{\rm CO}$ cancels out with the same term in Eq. \ref{tpscEq}.\\
In Eq. \ref{tpscEq}, we use an estimated preshock density $n_{H_{2},0}$ = 4$\cdot$10$^{2}$ cm$^{-3}$, v$_{\rm s}$ = 0.9 km s$^{-1}$ and assume that the equality in Eq. \ref{tpscEq} roughly holds. This implies that our mass accretion rate estimate here will be a lower limit. Putting all this together results in an estimated minimal mass accretion rate of 20 M$_{\odot}$ pc$^{-1}$ Myr$^{-1}$.

\end{appendix}

\end{document}